\documentclass{article}
\usepackage{marginnote}
\usepackage{url}
\usepackage{soul}
\usepackage[many]{tcolorbox} 
\usepackage[letterpaper,left=0.75in,right=2.25in,marginparwidth=1.5in]{geometry}

\definecolor{main}{HTML}{5989cf}    
\definecolor{sub}{HTML}{cde4ff}     

\tcbset{
    sharp corners,
    colback = white,
    before skip = 0.2cm,    
    after skip = 0.5cm      
}                           

\newtcolorbox{boxH}{
    colback = sub, 
    colframe = main, 
    boxrule = 0pt, 
    leftrule = 6pt 
}

\newtcolorbox{boxA}{
    fontupper = \bf,
    boxrule = 1.5pt,
    colframe = black 
}

\title{Patient-specific, mechanistic models of tumor growth incorporating artificial intelligence and big data}

\author{Guillermo Lorenzo$^{1,2}$, 
Syed Rakin Ahmed$^{3,4,5,6}$,  
David A. Hormuth, II$^{2,7}$, \and 
Brenna Vaughn$^8$, 
Jayashree Kalpathy-Cramer$^9$, 
Luis Solorio$^8$, \and 
Thomas E. Yankeelov$^{2,7,10-13}$, 
Hector Gomez$^{8,14,15}$}

\date{
$^1$Department of Civil Engineering and Architecture, University of Pavia, Pavia, Italy, 27100; email: guillermo.lorenzo@unipv.it\\ 
$^2$Oden Institute for Computational Engineering and Sciences, The University of Texas at Austin, Austin, TX, USA, 78712; email: guillermo.lorenzo@utexas.edu, david.hormuth@austin.utexas.edu\\
$^3$Athinoula A. Martinos Center for Biomedical Imaging, Department of Radiology, Massachusetts General Hospital, Boston, MA, USA, 02129; email: srahmed@mgh.harvard.edu\\
$^4$Harvard Graduate Program in Biophysics, Harvard Medical School, Harvard University, Cambridge, MA, USA, 02115; email: syedrakin\textunderscore ahmed@fas.harvard.edu\\
$^5$Massachusetts Institute of Technology, Cambridge, MA, USA, 02139; email: rakin@mit.edu \\
$^6$Geisel School of Medicine at Dartmouth, Dartmouth College, Hanover, NH, USA, 02139; email: syed.rakin.ahmed.med@dartmouth.edu\\
$^7$Livestrong Cancer Institutes, The University of Texas at Austin, Austin, TX, USA, 78712\\
$^8$Weldon School of Biomedical Engineering, Purdue University, West Lafayette, IN, USA, 47907; email: vaughn53@purdue.edu, lsolorio@purdue.edu, hectorgomez@purdue.edu\\
$^9$Department of Ophthalmology, University of Colorado Anschutz, Denver, Colorado, USA, 80045; email: jayashree.kalpathy-cramer@cuanschutz.edu\\
$^{10}$Department of Biomedical Engineering, The University of Texas at Austin, Austin, TX, USA, 78712; email: thomas.yankeelov@utexas.edu\\
$^{11}$Department of Oncology, The University of Texas at Austin, Austin, TX, USA, 78712; email: thomas.yankeelov@utexas.edu\\
$^{12}$Department of Imaging Physics, MD Anderson Cancer Center, Houston, TX, USA, 77030\\
$^{13}$Department of Diagnostic Medicine, The University of Texas at Austin, Austin, TX, USA 78712\\
$^{14}$School of Mechanical Engineering, Purdue University, West Lafayette, IN, USA, 47907\\
$^{15}$Purdue Center for Cancer Research, Purdue University, West Lafayette, IN, USA, 47907\\
}

\begin{document}
\maketitle
\begin{abstract}
Despite the remarkable advances in cancer diagnosis, treatment, and management that have occurred over the past decade, malignant tumors remain a major public health problem. Further progress in combating cancer may be enabled by personalizing the delivery of therapies according to the predicted response for each individual patient. The design of personalized therapies requires patient-specific information integrated into an appropriate mathematical model of tumor response. A fundamental barrier to realizing this paradigm is the current lack of a rigorous, yet practical, mathematical theory of tumor initiation, development, invasion, and response to therapy. In this review, we begin by providing an overview of different approaches to modeling tumor growth and treatment, including mechanistic as well as data-driven models based on ``big data" and artificial intelligence. Next, we present illustrative examples of mathematical models manifesting their utility and discussing the limitations of stand-alone mechanistic and data-driven models. We further discuss the potential of mechanistic models for not only predicting, but also optimizing response to therapy on a patient-specific basis. We then discuss current efforts and future possibilities to integrate mechanistic and data-driven models. We conclude by proposing five fundamental challenges that must be addressed to fully realize personalized care for cancer patients driven by computational models.
\end{abstract}


\tableofcontents

\section{INTRODUCTION}


Over the past three decades remarkable advances in cancer diagnosis, treatment, and management of malignant tumors have resulted in a 33\% overall reduction in cancer mortality \cite{siegel_cancer_2023}, yet the incidence of many cancers continues to increase and cancer remains a major public health problem worldwide. Shifting the clinical management of these diseases towards personalized therapies that account for the intra- and inter-tumor heterogeneity in treatment response could further improve patient outcomes.  However, it is not typically known \emph{a priori} how a patient’s tumor physiology, genotype, and phenotype will influence response to a particular therapeutic regimen.  Thus, the current clinical paradigm for personalized therapeutic regimens relies on \emph{post hoc} assessment of tumor response after some or all of the therapy has been delivered. Unfortunately, this approach potentially exposes patients to invasive biopsies, weeks or months of ineffective therapy, and off-target toxicities. Personalized therapies guided by a patient’s predicted outcome rather than their observed or assessed outcome could avoid these ineffective therapies and their side effects and dramatically improve patient outcomes \cite{yankeelov_toward_2015}.
To design personalized therapies guided by predicted patient response requires a deep understanding of the underlying biology embedded within a practical (i.e., actionable) mathematical theory capable of rigorously characterizing the dominant processes determining the individual’s response to a particular set of interventions. Without a mathematical theory rooted in known cancer biology, we are left with trial and error (albeit, perhaps, from well-informed large population-based studies) to select the optimal therapeutic regimen for a given patient. In fact, there has been enormous effort to develop mathematical descriptions of the disease from the genetic to tissue scales \cite{Yankeelov2016,bull_hallmarks_2022,Deisboeck2011} in the growing field of mathematical or computational oncology \cite{anderson_mathematical_2018,Rockne2019}. 

Mathematical models that attempt to incorporate the underlying mechanisms of tumor biology can provide readily interpretable solutions that enable the systematic investigation of alternative treatment strategies \emph{in silico} to estimate their efficacy, safety, and overall impact on tumor response, thereby helping the treating physicians to make well-informed (high-risk) decisions. To fully realize the potential of mechanistic mathematical models to guide care on a patient-specific basis, models need to be initialized and integrated with patient-specific data that is practically available in the routine clinical setting. While there is hardly a scarcity of mathematical models describing tumor initiation, growth, invasion, and response \cite{bull_hallmarks_2022,anderson_mathematical_2018}, a historical (and lingering) challenge has been that many of these models rely on parameters that are difficult or impossible to assign or calibrate based on readily-available data. Some of the barriers to progress have been the lack of quantitative biological assays for directly measuring model parameters or the necessity of highly-invasive (or impractical) means to measure quantities of interest \cite{Kazerouni2020}. Therefore, in the development of a clinically-relevant mechanistic modeling framework, it is important to determine the quantities of interest that can be measured at the spatial and temporal scales necessary to personalize the key biological components of the mechanistic model. Fortunately, in parallel to the advances of the clinical management of cancer, there has been substantial growth in the development of mathematical models that can be initialized and personalized \emph{via} patient-specific data \cite{Kazerouni2020,Baldock2013,hormuth_opportunities_2022,Enderling2019,jarrett_quantitative_2021,ocana-tienda_growth_2022} through either direct measurement or model calibration and therefore might serve as tools enabling personalized therapy.
\marginpar[AI: Artificial Intelligence]{AI: Artificial Intelligence}

On the opposite side of the computational spectrum from mechanism-based models, are the methods of artificial intelligence (AI) and “big data”. While the potential of AI to positively impact healthcare is difficult to deny, there are fundamental limitations of the AI-only approach that may restrict its ability to hasten the arrival of truly personalized medicine. In particular, the AI approach relies on data sets that may fail to capture unique individual patient conditions and are not truly representative of a diverse patient population \cite{shreve_artificial_2022}. These issues are exacerbated when trying to identify an optimal therapy for a particular patient’s tumor subtype and physiological conditions—all of which can exist outside of the available training data. However, AI may serve a complementary role with mechanistic modeling in realizing personalized medicine through, for example, relating large quantities of ‘omic’ data to mechanistic model parameters, reducing the computational burden, or parsing mechanistic model forecasts to select optimal therapies \cite{Rockne2019,baker_workshop_2019}. 

In this review we discuss topics from theory to application that may help realize personalized medicine in oncology. Clinically-relevant models of tumor growth have been and will continue to be driven by the available data. Thus, we first describe the available data across scales. Second, we introduce mathematical models based on biological mechanisms and data-driven models facilitated by AI. Third, we discuss how mechanistic models have the potential to identify optimal interventions and treatments. Fourth, we describe how mechanistic models and AI can be synergistically combined to advance the fields of personalized cancer management and treatment. 
Finally, we conclude by identifying five fundamental and currently unmet challenges to realizing truly personalized care for cancer patients. 


\section{DATA AVAILABILITY ACROSS SCALES}

\subsection{Microscopy approaches for informed diagnostics}

Traditional pathological examination of tissue samples has been performed under a brightfield microscope by highly trained laboratory personnel. However, a new era of digital pathology is evolving towards digitized whole-slide images where the whole prepared slide is scanned with an automated scanning microscope \cite{RN51}. The availability of these detail-rich digital files has facilitated the rise of computer aided/assisted diagnosis (CAD) which aims to increase speed, accuracy, and sensitivity in the diagnosis of breast cancer and other malignancies \cite{RN52}. CAD relies on the digitization of whole-slide images of prepared tissue sections that are usually stored in a high-resolution format. However, large files are challenging for centers to store and manage. These data can also be sensitive to center-based and user-based perturbations. Labor-intensive staining of the slides introduces variation between laboratories, creates heterogeneity in samples and can influence diagnostic model training and performance on unseen data known as domain shift \cite{RN53}. Domain shift represents a significant problem in the computer automated pathology realm \cite{RN54}. Core needle biopsy histology has shown promise in controlling domain shift but is still not ready for clinical application \cite{RN55}. Currently, histological analysis of tumor sections by pathology professionals remains the gold standard diagnostic technique for solid tumors \cite{RN52}. There remains a need for large, high-fidelity data sets from tissue samples to train cancer cell detection in samples for screening patients, monitoring disease progression, and interoperative tumor margin applications. Attenuated Total Reflection Fourier Transform Infrared (ATR-TFIR) spectroscopy represents a potential resource in training machine learning models to detect cancer cells. ATR-FTIR has shown promise as a high-throughput method of detecting biochemical changes in malignant cells \cite{RN56, RN57}. ATR-FTIR is a biochemical analysis method that can discriminate a sample’s unique spectra of lipids, proteins, carbohydrates, and nucleic acids. These spectra, a biochemical “fingerprint” of the sample, represent a large training dataset with potential to couple with machine learning to offer reliable quantitative information readily usable by computational models. 
\marginpar[CAD: Computer Aided/Assisted Diagnosis]{CAD: Computer Aided/Assisted Diagnosis\\}
\marginpar[ATR-TFIR: Attenuated Total Reflection Fourier Transform Infrared]{ATR-TFIR: Attenuated Total Reflection Fourier Transform Infrared\\}
\subsection{Medical imaging}\label{sec:medical-imaging}

Medical imaging is central to the clinical management of cancer patients as it is used for screening, diagnosis and staging, guiding interventions, assessing response, and long-term surveillance. It is widely available and a trusted data source capable of delivering longitudinal, non-invasive (with the exception of the administration of exogeneous contrast agents or radiotracers), and spatially resolved measurements of tumor biology. These characteristics make medical imaging an exceptional candidate for informing mechanistic tumor models \cite{Lorenzo2023}. 
\marginpar[MRI: Magnetic resonance imaging]{MRI: Magnetic resonance imaging\\}
\marginpar[PET: Positron Emission Tomography]{PET: Positron Emission Tomography}
Magnetic resonance imaging (MRI) \cite{Hormuth2019a} and positron emission tomography (PET) \cite{jin_positron_2022} have emerged as the primary modalities for informing patient-specific models of tumor growth and response at the tissue scale. In the routine clinical setting, MRI can achieve a spatial resolution of 1-2.5 mm isotropic with a temporal resolution of seconds to minutes, whereas PET may achieve a spatial resolution ranging from 2.5 to 5 mm isotropic with a temporal resolution of minutes to hours. MRI is capable of quantitatively characterizing anatomy and underlying physiology of the tissue of interest, while PET can report on a range of molecular and metabolic features \emph{via} a wide range of radiotracers. Since anatomical imaging provides information about the structure and extent of the tumor, it plays a crucial role in the assessment of treatment response across disease sites \cite{eisenhauer_new_2009,wen_updated_2010}. For mathematical modeling applications, anatomical images are essential for defining the computational domain, assigning boundary conditions, and identifying the extent of disease both before and during treatment. Physiological or molecular imaging techniques report on an array of parameters frequently incorporated into mechanism-based models including: a) cellularity, \emph{via} diffusion-weighted MRI (DW-MRI) \cite{padhani_diffusion-weighted_2009}; b) vascularity and perfusion, \emph{via} dynamic contrast-enhanced MRI (DCE-MRI) \cite{oconnor_dynamic_2011,quarles_imaging_2019}; c) hypoxia, \emph{via} 18F-fluoromisonidazole PET \cite{rajendran_hypoxia_2004}; and d) glucose metabolism, \emph{via} 18F-flourodeoxyglucose PET \cite{castell_quantitative_2008}. Mapping these tumor characteristics in three dimensions and across time facilitates patient-specific, tissue-scale modeling as the mechanism-based mathematical models frequently explicitly incorporate such quantities of interest. It is important to note that heterogeneity in these parameters can have great influence on patient response; for example, perfusion or tissue oxygenation are well-established indicators of the efficacy of systemic- and radio-therapy. A patient-specific model capable of predicting how these quantities change in space and time could provide guidance to many clinical management strategies across oncology \cite{vaupel_treatment_2001}. Therefore, accurate measurement and characterization of the evolution of these properties \emph{via} mechanistic mathematical modeling may be essential to optimizing treatment strategies and, subsequently, treatment outcomes on a patient-specific basis. 
\marginpar[DW-MRI: Diffusion weighted MRI probes water diffusion]{DW-MRI: Diffusion weighted MRI probes water diffusion\\}
\marginpar[DCE-MRI: Dynamic contrast-enhanced MRI measures tissue perfusion and permeability]{DCE-MRI: Dynamic contrast-enhanced MRI measures tissue perfusion and permeability\\}

\subsection{Informing on the totality of molecular processes}

Tumors are comprised of dynamic, heterogenous cell populations which change as the cancer progresses \cite{RN1}. These changes not only allow cells to seed and colonize new tissues, but they also alter the response of tumor cells to therapeutics. Precision analysis of these mixed cell populations poses a substantial hurdle as bulk sequencing of cells (i.e., all of the cells simultaneously) results in population averaging \cite{RN2}, and can limit insight into how the mixed cell populations interact to facilitate metastasis and drug resistance. Single cell methods of characterizing individual tumor cells overcome this challenge and provide rich datasets on the cell’s genome, epigenome, proteome, or transcriptome \cite{RN3}. Single cell genomic sequencing approaches such as the widely-adopted scRNA-seq can provide information about tumor cell population heterogeneity and how cell populations change with growth and metastasis \cite{RN4}. Multiomics approaches that allow for the profiling of multiple “omes” in an individual cell are rising in popularity \cite{RN5}. As the amount of multimodal data extracted from singular cells is increasing, cellular labeling methods such as molecular barcoding become increasingly important to tie data back to its source \cite{RN5,howland2023cellular}. Despite the increasing complexity of data able to be extracted from each cell, omics data are inherently static as the cell must be lysed \cite{RN4, RN6}. Therefore, data that could have been gleaned from dynamic processes suffers. Deployment of machine learning algorithms on omics datasets generated from patient samples as the disease progresses represents a significant opportunity for patient-specific medicine. 
\marginpar[scRNA-seq: Single‐cell RNA sequencing, a powerful tool for studying gene expression at the level of individual cells]{scRNA-seq: Single‐cell RNA sequencing, a powerful tool for studying gene expression at the level of individual cells\\}


\section{MATHEMATICAL MODELS}

\subsection{Mechanistic and data-driven: Two types of models to fight cancer}


Finding a mathematical model that faithfully reproduces the spatiotemporal dynamics of a tumor may be considered the ultimate goal of tumor modeling. However, there are numerous examples in tumor modeling (and other fields of science) showing that, even if that ultimate goal is not achieved, careful development of parsimonious models can provide remarkable insight and useful approximations \cite{box1979robustness}.

Mechanistic models are based on first principles and hypotheses generated from observations. In cancer modeling, first principles include physics principles like conservation of mass and biological principles like the concept of homeostasis. Observations include medical images, pathological information obtained from microscopy, genomic and proteomic information, and clinical data. Data-driven models are consistent with observations, but cannot be derived from first principles. The purpose of data-driven models is to enable prediction. Mechanistic models are used to increase our understanding of a phenomenon, but they can also be used for prediction after calibration. The distinction between mechanistic and data-driven models goes back many years \cite{neyman1939new}. Historically, scientists have favored mechanistic over data-driven or phenomenological models. However, recent literature has seen a more vigorous debate. The unprecedented growth of data-driven models is fueled by AI, and our increasing ability to collect, fit, and extrapolate large amounts of data. This growth is driven by advances in hardware and software, including tensor processing units (i.e., a circuit specifically designed to accelerate AI workloads) and machine learning algorithms.



\subsection{Data-driven models} 
\subsubsection{The AI paradigm}
AI, and its most recent and exciting subset, deep learning, has permeated virtually every aspect of modern life. Medicine has been no exception: Based on PubMed data, the number of publications that use the terms ``deep learning" or ``neural network" has increased by a factor of $\sim$10 in the last 5 years. The most common application of AI is image classification. Therefore, it is not surprising that quantitative techniques for deep learning-based image analysis have gained momentum in medicine. The reasons for the explosion of AI in medical imaging are multifold, but two facets deserve especial attention: a pressing and consistent clinical need, and the convergence of computational innovations with interdisciplinary collaborations. These facets are particularly well highlighted in the context of the cancer continuum, from diagnosis and detection to treatment and management of malignant lesions. Given the heterogeneity and dynamic nature of cancers, medical advances have focused on earlier screening and detection, and preferential removal of lesions using a subset or a combination of surgery, chemotherapy, radiation therapy, immunotherapy, and/or targeted therapy, to improve survival. Here, we describe the two arms of the tumor progression continuum where AI has shown significant success, namely, diagnosis and treatment assessment, as well as a more emerging application intended to predict diagnostic endpoints.

\subsubsection{AI for cancer diagnosis}
Initial cancer detection and staging today employs a variety of noninvasive or minimally-invasive imaging technologies as described in Sect.~\ref{sec:medical-imaging}. The diagnostic gold standard for most cancers, however, remains the biopsy. A biopsy presents a non-negligible degree of invasiveness to the patient regardless of cancer type, as sample tissue must be extracted from the body for grading \cite{lone2022liquid}. Given the minimally invasive nature of the aforementioned imaging modalities already a part of the clinical workflow, there has been a proliferation of AI-based algorithms, many of which claim to interpret imaging scans at the level of expert clinicians. In  breast cancer, a quantitative image analysis technique for detection of Human epidermal growth factor receptor 2 (HER2) has received FDA approval \cite{her2ai,baxi2022digital,vandenberghe2017relevance}. The goal of a HER2-directed image analysis platform is to detect and quantify HER2 membranous IHC staining, providing critical information to guide treatment. 
In prostate cancer, Paige \cite{raciti2022clinical} is the first AI-based pathology product to receive FDA approval for {\em in vitro} diagnostic use to detect cancer in prostate biopsies. In brain cancer, the release of annotated, open-source datasets, such as the Multimodal Brain Tumor Segmentation Challenge (BraTS) \cite{Menze2015,baid2021rsna,bakas2017advancing,bakas2017segmentation,bakas2017segmentation2} has spearheaded the generation of AI models for tumor segmentation, which is essential for planning surgical and radiation treatments. Segmentation of primary brain tumors has been often performed with 3D U-Nets and their variations \cite{DBLP:journals/corr/RonnebergerFB15}. Notably, the winners of the 2020 BRATS challenge used an architecture named ``No New-Net" to emphasize that a standard U-Net coupled with thorough training and optimization continues to achieve state-of-the-art results \cite{isensee2021nnu}.

\marginpar[HER2: Human epidermal growth factor receptor 2; overexpression is associated with breast cancer invasiveness]{HER2: Human epidermal growth factor receptor 2; overexpression is associated with breast cancer invasiveness\\}
\marginpar[BraTS: Multimodal Brain Tumor Segmentation Challenge]{BraTS: Multimodal Brain Tumor Segmentation Challenge\\}

\vspace{1cm}

\begin{boxH}
DEEP LEARNING SEGMENTATION WITH nnU-Net\\
The nnU-Net is a segmentation method that utilizes deep learning to automatically adapt to new tasks by configuring various components such as preprocessing, network architecture, training, and post-processing. The process involves incorporating fixed parameters, interdependent rules, and empirical decisions to make key design choices. By initially extracting the ``dataset fingerprint" and applying heuristic rules, nnU-Net generates three distinct configurations of U-Net: a 2D U-Net, a 3D U-Net operating at the original image resolution, and a 3D U-Net cascade. In the cascade, the first U-Net operates on down-sampled images, and the second U-Net is trained to enhance the segmentation maps produced by the first U-Net at full resolution \cite{isensee2021nnu}.
\end{boxH}

\subsubsection{AI for cancer progression assessment}
AI approaches aimed at assessing tumor progression can be categorized into radiomics and deep learning. Radiomics involves the extraction of quantitative imaging features and subsequent analysis using statistical and machine learning methods, while deep learning focuses on training deep neural networks to automatically learn patterns and relationships directly from raw medical image data \cite{wagner2021radiomics}. 
\marginpar[Radiomics: A data-driven approach for extracting features from imaging data]{Radiomics: A data-driven approach for extracting features from imaging data\\}
Current radiomics and deep learning applications can be delineated into three overarching areas: a) Preprocessing, which consists of approaches to automate image acquisition and spatiotemporal registration \cite{zhu2018image,fu2020deep}; b) Analysis, which includes anatomical segmentation, volumetric quantification, extraction of parameter maps from diffusion or perfusion imaging, and groupwise population analyses \cite{isensee2021nnu,10.1093/neuonc/noz106}; and c) Interpretation, which can be framed as an image classification process that includes identifying tumor subtypes, radiogenomics, response assessment, and survival prediction \cite{Baid2020,Han2020}. The longitudinal measurement of lesion burden is an important aspect of tumor progression assessment that is well-suited for AI. Even though volumetric measurement is ideal to assess lesion burden, proxy measures, such as RANO (Response Assessment in Neuro-Oncology) for gliomas are often used. The tool AutoRANO, which uses the outputs of a segmentation model that can run on post-operative images, aims at automating RANO measurements \cite{10.1093/neuonc/noz106}. Other tumor types have also seen promising advances in this area \cite{peng2022deep}. 
\marginpar[RANO: Response Assessment in Neuro-oncology]{RANO: Response Assessment in Neuro-oncology\\}
Another aspect of tumor progression assessment where AI has shown promising results is the automatic delineation of tumor subtypes and molecular markers from medical images. The correlation between imaging features and specific gene expressions of tumors is termed radiogenomics. \emph{A priori} knowledge of the mutational status of known oncogenic drivers together with radiographic suspicion of a neoplasm may favor prompt intervention, as well as targeted therapy. Some of the most promising advances in radiogenomics have enabled delineation of the Isocitrate Dehydrogenase (IDH) mutation status in gliomas \cite{chang2018residual}, 
and the Epidermal Growth Factor Receptor (EGFR) status in non-small cell lung cancer \cite{coudray2018classification}.
\marginpar[Radiogenomics: Methodology that combines radiomics features from medical images with genomic data from high-throughput sequencing]{Radiogenomics: Methodology that combines radiomics features from medical images with genomic data from high-throughput sequencing\\}

Finally, survival analysis, which is used in cohort and other longitudinal studies to estimate the time it takes for a specific event to happen, has incorporated radiomics, deep learning, and a combined utilization of both approaches in various tumor types. In the case of gliomas, for example, reference \cite{Baid2020} put forth a three-step framework for predicting overall survival, which involved segmenting data, extracting radiomic features, and utilizing a survival prediction model. The aim was to classify patients into three survival groups (short-term, mid-term, and long-term) and provide predictions for their overall survival \cite{patel2021opportunities}.



\subsubsection{AI for outcome prediction}
\begin{figure}
\centering
\includegraphics[width=\linewidth]{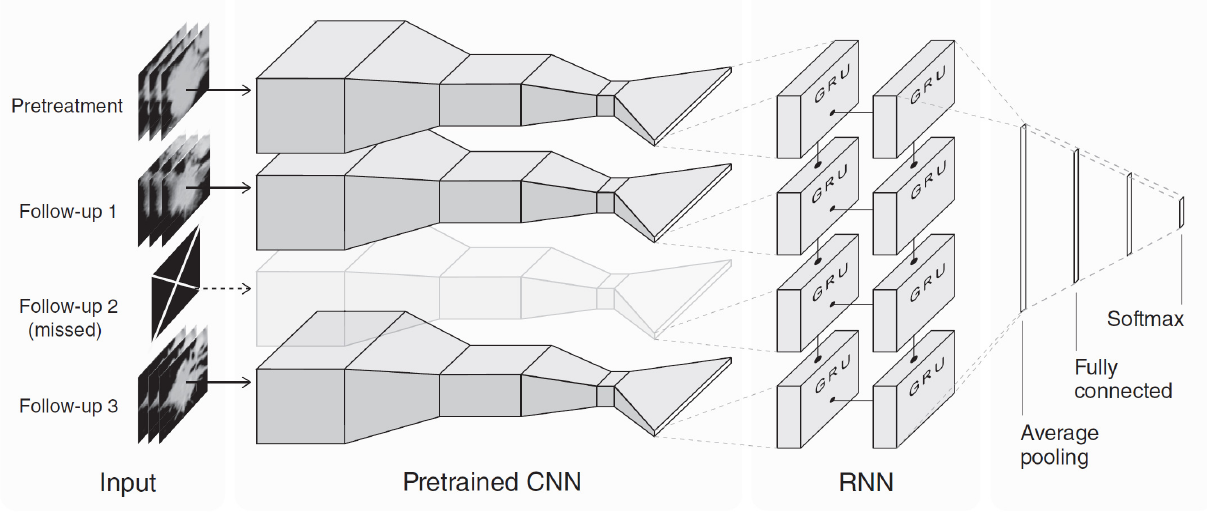}
\caption{\textbf{Data-driven model predicting outcome in lung cancer patients.} The input to the model is a sequence of longitudinal CT images from an individual patient that includes a pretreatment CT and several follow-up images. The network architecture is composed of a ResNet convolutional neural network (CNN) that was pretrained on the ImageNet database. Each input CT, corresponding to data at one timepoint, is fed into a CNN. The pretrained CNN is followed by a recurrent neural network (RNN) with gated recurrent units (GRU). The network is able to handle missing scans by masking the output of the CNN to skip the timepoint. After the GRU, the network uses averaging and fully-connected layers to prevent overfitting. The final softmax layer provides a binary classification output. Reproduced with permission from \cite{xu2019deep}.}
\label{fig_lung}
\end{figure}

While tumors are dynamic biological systems, influenced by their microenvironment and other factors, most AI algorithms have concentrated on developing imaging markers for a single time point. Contemporary deep learning techniques, such as recurrent neural networks (RNN), which have been successfully employed in video classification and natural language processing, have the potential to incorporate longitudinal data \cite{gregor2015draw,donahue2015long}. However, these advanced computational approaches have been relatively underutilized in the field of radiology. Recently, an algorithm combining convolutional neural networks (CNNs) and RNNs has shown success predicting the survival and pathologic response of patients with non-small cell lung cancer by integrating pretreatment and follow-up CT images \cite{xu2019deep}. In particular, this algorithm predicted two-year overall survival with AUC = 0.74. Fig.~\ref{fig_lung} shows the network architecture, where the pretrained CNN is a ResNet. While acquiring enough longitudinal data to train this type of networks is very challenging (n=107 patients were required for training) in cancer, this research highlights the potential of AI for outcome prediction.
\marginpar[CNN: Convolutional Neural Network]{CNN: Convolutional Neural Network\\}
\marginpar[RNN: Recurrent Neural Network]{RNN: Recurrent Neural Network\\}

\subsubsection{Fundamental Limitations}

The explosion of AI in elucidating tumor biology, while promising, has brought forth a number of key concerns. These limitations include dataset level limitations such as data variability, noise, poor quality and limited availability of large amounts of data, as well as model brittleness and lack of interpretability or explainability of the model \cite{bansal2020sam}.

Most current AI models continue to be highly dependent on the training dataset. These datasets are often limited in size and are frequently homogeneous. Data heterogeneity continues to be a key hindrance to the generalizability of deep learning models. Within real world clinical oncology data, this heterogeneity can come in several forms, such as different image capture devices, patient populations or inter-tumor heterogeneity. There is a paucity of current work towards the creation of clinical-domain-specific yet generalizable AI models. AI models also frequently struggle with cases that deviate significantly from the patterns seen in the training data, a phenomenon described as out-of-distribution data \cite{alcorn2019strike,eykholt2018robust}. Modeling tumor growth often requires approaches that are patient specific, and an AI-only solution is likely insufficient to capture the high patient-level variability in tumor growth.

Additionally, deep learning models are frequently considered as ``black boxes” owing to the minimal transparency in how these models arrive at predictions \cite{linardatos2020explainable}. This makes translating deep learning model predictions to clinical settings difficult \cite{vinuesa2021interpretable}. These models frequently have complex architectures and operate \emph{via} multiple layers of interconnected neurons, which makes understanding the reasoning behind their predictions difficult. This limits the ability to gain insights into the specific mechanistic patterns that drive tumor growth and the molecular makeup of the tumor micro-environment, and limits the reliability of AI-generated results.


\subsection{Mechanistic models}

\marginpar[ADC: Apparent diffusion coefficient; measure of water diffusion from DW-MRI]{ADC: Apparent diffusion coefficient; measure of water diffusion from DW-MRI\\}
\marginpar[T1-W MRI: T1-weighted MRI anatomical image]{T1-W MRI: T1-weighted MRI anatomical image\\}
\marginpar[T2-W MRI: T2-weighted MRI anatomical image]{T2-W MRI: T2-weighted MRI anatomical image}

The main mechanisms underlying the development of a tumor and the effect of treatments are described with mathematical models, which are usually written in terms of time-resolved ordinary differential equations (ODEs) or spatiotemporally-resolved partial differential equations (PDEs) and agent-based models. In general, the choice of each formulation paradigm is subject to the scale of the tumor-specific phenomena to be captured by the model (e.g., cell-cell interactions in the tumor microenvironment, chemotherapeutic response at organ scale, metastasis over the body), as well as the available data types, and their level of spatiotemporal resolution (e.g., cellular versus organ scale, and daily versus yearly measurements) \cite {Kazerouni2020, Metzcar2019, Lorenzo2023}. Here, we focus on models based on ODEs and PDEs, but agent-based models also have a rich literature and are important to investigate the spatiotemporal evolution of a tumor and its microenvironment at cellular level \cite{Metzcar2019}.

\subsubsection{Mechanistic models based on ODEs}

ODE models have been broadly leveraged to describe the temporal dynamics of tumor burden and its response to treatment both in preclinical and clinical scenarios. These models are posed in terms of
variables that can be repeatedly measured over time during the
experimental protocol or the clinical management of the tumor, such that
these longitudinal datasets can be used to inform the model. There are
three common choices for these variables. First, tumor volume is an
established metric for monitoring therapeutic response that can be
readily determined, for example, using anatomic imaging data in
(pre)clinical settings (e.g., X-ray CT, and T1- and T2-weighted MRI) and also calipers in animal models \cite{Lima2022, Zahid2021, Slavkova2023}. Second, the tumor cell count is a common metric of tumor malignancy and therapeutic response
that can be measured, for instance, using \emph{in vivo} ADC maps
derived from DWI-MRI and automated microscopy techniques during \emph{in
vitro} experiments \cite{Kazerouni2020, Yang2022}. Finally, some tumor-specific biomarkers can serve as surrogates for tumor growth and response to treatment, such as the blood serum levels of the Prostate-Specific Antigen (PSA) in the clinical management of prostate cancer \cite{BradyNicholls2020, Strobl2021, Lorenzo2022}.

\begin{figure*}[!t]
\centering
\includegraphics[width=\linewidth]{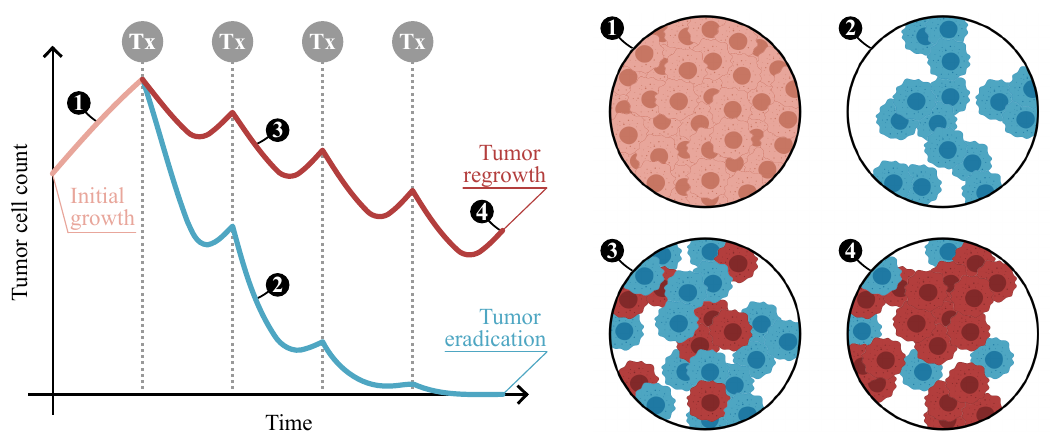}
\caption{\textbf{ODE modeling of tumor growth and treatment response.}  
(Left) Example of the outcome of an ODE model describing the temporal dynamics of tumor cell count under a certain cytotoxic therapy. Vertical dotted lines indicate the times of treatment (Tx) delivery. An initial period of untreated growth is represented in light orange, and two alternative scenarios of therapeutic response are considered after treatment onset. The blue curve represents a situation in which the therapeutic regimen effectively kills all tumor cells. The red curve corresponds to a scenario in which 
the tumor resumes growth after the last therapy dose. This situation may be motivated by different causes, such as the existence of a subset of treatment-resistant cells in the original tumor, the development of chemoresistance or an inadequate treatment regimen for this specific tumor. (Right) Snapshots numbered 1 to 4 provide insight into the tumor cells status in different situations. In these snapshots, light orange cells represent untreated cells, blue cells are effectively treated and will ultimately die, and red cells survive to the chosen therapy and continue proliferating. This figure was partially created using BioRender.
}
\label{fig_ode}
\end{figure*}
\marginpar[left]{PSA: Prostate-specific antigen, a biomarker of prostate cancer}

The formulation of ODE models usually features a term describing tumor
growth and one or more terms characterizing the effects of treatments on
tumor dynamics \cite{Benzekry2014, Yin2019, Lima2022}. For
instance, radiotherapy is usually represented as an instantaneous
reduction of the tumor burden on the dates of each radiation dose,
although this approach can also be extended to a hybrid formulation with
a continuous term accounting for the delay in tumor cell death since the
exposure to radiation \cite{Lorenzo2022, Zahid2021}. In general, drug-based therapies use a hybrid formulation to represent the treatment: the discrete component represents the delivery of drug doses while continuous terms may account for a diverse host of mechanisms, such as a dose-dependent death rate, early and late effects of the
drugs, drug interactions, and the time-resolved change of drug concentration during treatment \cite{Lima2022, BradyNicholls2020, Yang2022}. 
For example, an ODE model to describe the temporal dynamics of tumor growth and response to chemotherapy in terms of the total tumor cell count ($N$) can be written as 
\begin{equation}\label{odeq}
    \frac{dN}{dt}=\rho N\left(1-\frac{N}{K}\right) - \alpha N U_0 \sum_i e^{-\beta (t-t_i)}\mathcal{H}\left(t-t_i\right).
\end{equation}
The first term on the right-hand side of Eq.~\ref{odeq} represents tumor growth with a logistic term, in which $\rho$ is the net tumor cell proliferation rate and $K$ is the tissue carrying capacity (i.e., the maximally admissible tumor cell burden). The second term on the right-hand side 
formulates the tumor cell-killing effect of a drug, where $\alpha$ is the rate of drug-induced tumor cell death, $U_0$ is the delivered drug dose, $\beta$ is the drug's decay rate, $t_i$ are the times of drug delivery, and $\mathcal{H}\left(t-t_i\right)$ is a Heaviside function activated at the treatment times. Figure~\ref{fig_ode} illustrates the result of an ODE model of tumor growth and response to cytotoxic therapy (e.g., chemotherapy, radiotherapy).

ODE models of tumor growth and therapeutic
response can also be extended to a multicompartment formulation that can
accommodate, for example, multiple cell types \cite{Strobl2021, BradyNicholls2020}, different therapeutic responses \cite{Yang2022,
Lorenzo2022}, several
imaging-identified intratumoral regions with prognostic significance
(e.g., tumor habitats with distinct cellularity and vascularization)
\cite{Slavkova2023}, or other variables affecting tumor dynamics and
eligible as treatment targets (e.g., tumor vascularity, reactive oxygen
species) \cite{Lima2022, Benzekry2014}. Consequently, these
multicompartment models consist of a set of ODEs informed by one or more
of the aforementioned time-resolved data types.

The main advantage of ODE models is that they require minimal computational resources, which facilitates the execution of multiple simulations in short times. This advantage is pivotal to enable a fast computational evaluation of multiple therapeutic regimens and find an optimal treatment for each individual \cite{Lima2022,Yang2022, BradyNicholls2020}. Additionally, the reduced computational cost of ODE models also facilitates the quantification and propagation of uncertainties in the data to the ensuing forecasts of tumor growth and therapeutic response \cite{Lima2022}. Importantly, uncertainty quantification is key to derive clinically-actionable modeling results supporting clinical decision-making, such as probabilistic risks and survival odds \cite{BradyNicholls2020, Zahid2021}. However, the main limitation of ODE models is their inherent lack of spatial resolution, which limits their ability to capture intratumor heterogeneity and its driving role in tumor development and therapeutic response.

\subsubsection{Mechanistic models based on PDEs}

\begin{figure*}[!t]
\centering
\includegraphics[width=\linewidth]{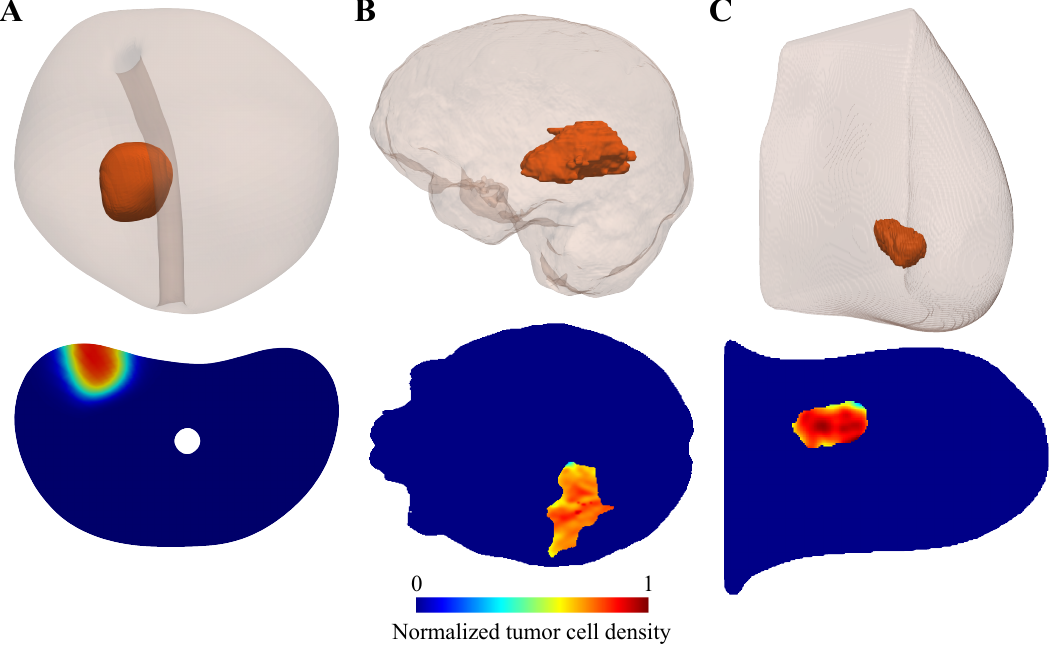}
\caption{\textbf{PDE modeling of tumor growth and treatment response.}  
This figure illustrates the outcome of three patient-specific PDE models that have been developed to predict prostate cancer growth during active surveillance (panel A) \cite{Lorenzo2022b}, response of high-grade glioma to chemoradiation (panel B) \cite{Hormuth2021}, and breast cancer response to neaodjuvant therapy (panel C; images courtesy of Dr. Chengyue Wu) \cite{Wu2022b}.
Each panel shows a 3D visualization of the patient's tumor within the affected organ on top, along with an axial section showing the map of normalized tumor cell density calculated with the PDE model.
}
\label{fig_pde}
\end{figure*}

PDE models constitute a natural extension of ODE models to accommodate
spatially-resolved mechanisms at tissue and organ scale in both
preclinical and clinical scenarios (see Fig.~\ref{fig_pde}). There are three main paradigms to
describe tumor growth and therapeutic response with PDE models. First,
advection-diffusion-reaction models are usually posed in terms of tumor
cell density \cite{ Wu2022b, Hormuth2021, Lipkova2019, Corwin2013,  Wong2016, Angeli2018} or tumor cell
volumetric fraction \cite{Urcun2021,
Kremheller2019}. The formulation of these models is obtained by
combining two types of phenomena: (i) a mobility mechanism, which is
generally represented by a flux featuring diffusive and advective
processes within an elastic \cite{Wu2022b, Hormuth2021,
Lipkova2019, Corwin2013} or poroelastic medium \cite{Angeli2018,
 Urcun2021, Kremheller2019},
and (ii) a collection of reaction terms representing local tumor cell
mechanisms, such as proliferation, death, metabolism, therapeutic
response, and phenotypic change of tumor cells \cite{Yin2019,
Lima2017, Lorenzo2023}. Second, phase-field models are governed by the
minimization of a functional featuring all types of energy interacting
in the tumor growth problem (e.g., internal energy, kinetic energy,
interfacial energy between tumor and healthy tissues, elastic energy),
wherein the tumor is represented by a variable identifying the
regions occupied by cancerous tissue \cite{Lorenzo2019, Colli2021, Fritz2021,
Lima2017, Xu2020, Wise2008}. The resulting PDE
model is completed with a collection of reaction terms as those
mentioned above, and may also feature advection \cite{Fritz2021, Wise2008}. Third, biomechanical models describe
tumor growth in terms of the nonlinear volumetric deformation of the
host tissue by leveraging a combination of poroelasticity and
multiplicative splitting to separate the elastic and growth-dependent
components of the deformation \cite{blanco2023mechanotransduction,Stylianopoulos2013, Vavourakis2018}. In these models, the constitutive definition of the main stretches in the growth deformation
gradient introduces the description of tumor growth and therapeutic
response using reaction-based formulations.

According to the description of the three types of PDE models provided
above, their formulation can accommodate the time-resolved terms
appearing on the right-hand side of ODE models into reaction terms with
a spatiotemporal definition. This feature enables the representation of
intratumoral heterogeneities, for example, using proliferation maps
calibrated from longitudinal imaging data \cite{Wu2022b,
Hormuth2021, Wong2016}, adjusting growth in terms of the
local availability of nutrients or the local vascularization
\cite{Roque2018, Angeli2018, Lorenzo2019,
Colli2021, Hormuth2019, Urcun2021}, and characterizing therapeutic response based on the local
tumor burden \cite{Yin2019, Corwin2013,  Wu2022b, Hormuth2021}, nutrient availability \cite{Angeli2018,
Rockne2015}, tumor-supporting vasculature \cite{Hormuth2020, Hormuth2021,
Swanson2011}, and drug concentration in the tissue \cite{Wu2022b, Vavourakis2018}. The description of these
phenomena driving tumor growth can be further enriched by extending a
baseline PDE model to include other key spatiotemporal mechanisms
represented by additional PDEs, such as the effect of tumor-driven
mechanical deformation of the host tissue and the effect of mechanical stress on tumor dynamics \cite{Hormuth2021,
 Wu2022b, Lorenzo2019, 
Stylianopoulos2013, Angeli2018, Wong2016}, nutrient and drug transport in the tissue
\cite{Vavourakis2018, Lorenzo2019, Fritz2021, Angeli2018}, the production of key substances (e.g., angiogenic
factors, matrix-degrading enzymes or biomarkers) \cite{Lorenzo2019,
Swanson2011, Wise2008, Vavourakis2018, Fritz2021}, blood flow
in the native vascular network \cite{Xu2020, Fritz2021, Kremheller2019},
and the development of angiogenesis \cite{Xu2020, Kremheller2019, Hormuth2019}. Similar to ODE models, PDE
models can also adopt a multicompartmental formulation to accommodate
several tumor cell types with different phenotypes or therapeutic
response (e.g., normoxic, hypoxic, and necrotic cells; sensitive and
resistant cells to a given treatment; surviving and irreversibly-damaged
cells after a therapeutic dose) \cite{Swanson2011, Wise2008, Roque2018, Fritz2021}, various non-tumor cell
and tissue types (e.g., healthy cells, vascular tissue,
extracellular matrix) \cite{Hormuth2019, Kremheller2019,
 Xu2020,
Wise2008}, as well as multiple solid and liquid phases included in the
problem (e.g., interstitial water and extracellular matrix)
\cite{Kremheller2019, Stylianopoulos2013,  Urcun2021,
Angeli2018, Fritz2021, Wise2008}.

To illustrate the formulation of PDE-based models of tumor growth and treatment response discussed herein, let us consider the extension of the ODE model in Eq.~\ref{odeq} to a spatiotemporal framework given by
\begin{equation}\label{pdeq}
    \frac{\partial N}{\partial t} = \nabla\cdot\left(\mathbf{D} \nabla N\right) + \rho N \left(1-\frac{N}{K}\right) - \alpha N U_0 \sum_i e^{-\beta (t-t_i)}\mathcal{H}\left(t-t_i\right).
\end{equation}
In Eq.~\ref{pdeq}, the first term on the right-hand represents the mobility of the tumor cell density $N(\mathbf{x},t)$ with a diffusion process governed by the tumor cell diffusivity tensor $\mathbf{D}(\mathbf{x},t)$, which can vary spatiotemporally (e.g., due to tumor-induced tissue deformation) or represent preferential directions of growth. The second and third terms on the right-hand side of Eq.~\ref{pdeq} are reaction terms matching those in Eq.~\ref{odeq}.
However, now parameters governing tumor cell proliferation and drug effects may adopt a spatiotemporal definition (e.g., $\rho(\mathbf{x},t)$, $\alpha(\mathbf{x},t)$).

PDE models are primarily informed by imaging data that, if collected at several time points, provide a spatiotemporal characterization of the tumor dynamics enabling model initialization, calibration, and regular update. Anatomic imaging data types, such as CT, T1-W MRI, and T2-W MRI (with and without contrast), have been extensively used to define a computational model of the 3D geometry of the tumor and its host organ from their corresponding imaging-based segmentations \cite{ Lorenzo2019, Lipkova2019, Wu2022b, Hormuth2021, Angeli2018}. These
imaging data can also be employed to define spatial maps of mechanical
properties for the different tissues or anatomic zones in the host organ
\cite{ Wu2022b, Hormuth2021, Lorenzo2019}. Additionally, anatomic
imaging data have been proposed to define tumor cell density maps, for
example, by combining T1-W and T2-W MRI data \cite{Corwin2013} or using pre-contrast and post-contrast CT data
\cite{Wong2016}. Quantitative imaging data types can
also provide measurements of tumor morphology, but these imaging types
can further probe tumor biology properties that are not accessible with
anatomic imaging. For example, ADC maps derived from DW-MRI have been
used to define tumor cell density maps \cite{Wu2022b,
Hormuth2021, Lorenzo2022b}, while DTI constitutes the main data
type to inform preferential directions of growth \cite{Gholami2016, Angeli2018}. DCE-MRI
can inform the vascular compartments of the model (e.g., the main vascular network in the host tissue or tumor-supporting vasculature formed due to angiogenesis) \cite{Hormuth2019, Wu2022} as well as local perfusion, which
can be leveraged as a surrogate for the relative availability of therapeutic agents and nutrients \cite{Wu2022b}. Additionally, DCE-MRI has also been proposed to perform tumor segmentation \cite{Hormuth2019, Lima2017, Wu2022b} and to define tumor cell densities of multiple tumor cell phenotypes according to nutrient availability \cite{Roque2018}. PET data can provide a host of biological information to inform PDE models depending on the specificity
of the radiotracer, such as tumor cell density maps \cite{Lipkova2019},
the local level of hypoxia to inform tumor cell radiosensitivity
\cite{Rockne2015} and the tumor metabolic activity to inform
proliferation rates \cite{Wong2016}. In preclinical settings,
while MRI and PET data can be collected from animal models
\cite{Hormuth2019, Judenhofer2013, Hormuth2019a},
photoacoustic and microscopy imaging techniques also enable to collect
spatially-resolved data from cell cultures  and tissue
samples \cite{Kazerouni2020, Xu2020, Urcun2021}. Moreover, the spatial integration of the
fields representing the location and amount of cellular types and the concentration of key substances over the host organ anatomy can provide scalar values comparable to common clinical measurements and key problem-specific quantities of interest (e.g., tumor volume, global cell
counts, blood levels of a biomarker like PSA, global amount of a nutrient or a growth factor or vascularity) \cite{ Lorenzo2019,
Lorenzo2022b, Xu2020, Wu2022b,
Hormuth2021}. Hence, these scalar measurements, which traditionally
inform ODE models, can also be used to constrain PDE
models as well.
\marginpar[left]{DTI: Diffusion tensor imaging; an imaging technique that measures anisotropic diffusion of water in tissue}

Therefore, the main advantage of PDE models is their ability to provide
a robust and flexible description of tumor growth and treatment response
at tissue and organ scale over timeframes ranging from weeks to years.
Nevertheless, their mathematical formulation also introduces two main
limitations. First, their numerical solution demands more computational
resources. 
Hence, the comparatively larger computational cost of PDE models may require ancillary methods to accelerate their simulation in applications requiring an elevated number of model simulations, such as treatment optimization or uncertainty quantification \cite{Lima2017, Mascheroni2021, Lipkova2019, Lorenzo2023}. The usual approaches to overcome this limitation include the development of fast and problem-specific solvers, as well as the deployment of reduced order modeling or scientific machine learning techniques \cite{Lorenzo2023, Lipkova2019, Alber2019, Viguerie2022}.
Second, PDE models that operate at tissue scale 
cannot resolve mechanisms specifically happening at cellular scale (e.g., cell-cell-interactions, asymmetric cell division, epithelial to mesenchymal transition, multiple phenotypes evolving with the local microenvironment and applied therapy). To address this limitation, hybrid multiscale models can introduce discrete agents in PDE models to characterize some of these cell-scale mechanisms at tissue scale (e.g., tip endothelial cells driving angiogenesis, epithelial to mesenchymal transition, taxis
phenomena); see \cite{ Xu2020, Vavourakis2018, Deisboeck2011, Yankeelov2016}.

\subsubsection{Limitations}
Beyond the limitations discussed thus far, the mathematical models described in this section are inherently constrained by the hypotheses underlying the formulation of the mechanisms participating in the model. These assumptions ultimately govern the number of parameters, variables, and equations constituting the model formulation and characterizing its complexity. Ideally, the complexity of ODE models and PDE models should be balanced with the spatiotemporal availability of data required to robustly calibrate them and render meaningful and accurate predictions. Model selection techniques can identify the optimal model formulation among a pool of candidates with increasing model complexity \cite{Hormuth2021, Lorenzo2023, Lima2017}. Preliminary models to describe a specific
type of tumor growth or therapeutic response usually start with a simple formulation capturing essential mechanisms (e.g., tumor cell mobility and net proliferation). The analysis and comparison of model simulations and available datasets can then reveal potential mechanisms that play a driving role in tumor growth and treatment action. Hence, the model can be extended to include these mechanisms, provided that there exists sufficient data to still perform a robust calibration of the new model. This path towards mechanistic model discovery has paved the way to
several model hierarchies, which provide an increasingly deeper insight in the biophysical processes underlying tumor growth and therapeutic response \cite{Lorenzo2023, Lima2017, Hormuth2021}.  Additionally, the recent application of machine learning techniques to model discovery has resulted in data-driven methods capable of identifying the mathematical formulation of ODE and PDE models \cite{Nardini2020, Brummer2023}, which show promise to identify hidden mechanisms underlying tumor and therapeutic dynamics as well as to improve model selection.

\section{MECHANISTIC MODELS ALLOW FOR OPTIMIZING INTERVENTIONS}

Multiple patient-specific mechanistic models have been shown to enable the personalized forecast of tumor growth and therapeutic response. Thus, these models provide an practical \emph{in silico} framework to investigate alternative monitoring and treatment strategies to those prescribed initially to each individual patient \cite{Lorenzo2023, Lipkova2019, Lorenzo2022, BradyNicholls2020}. In particular, these models could be exploited to perform an n=1 clinical study, where the outcomes of a range of clinical interventions are quantitatively assessed using computer simulations of the personalized model to rigorously and systematically optimize the monitoring and treatment of each patient's tumor. In this section, we discuss two approaches that have shown promise to achieve these
challenging goals: optimal control theory (OCT) and digital twins.

\subsection{Optimal control theory}

OCT is an established framework to optimize the state of a dynamical system through the action of external forces, which are systematically adjusted to achieve a particular goal of interest \cite{Lenhart2007, Lorenzo2023, Schattler2016}. The dynamics of the system is described by means of a mechanistic model, whose main variables are termed \emph{state variables,} while the external forces are termed \emph{controls} or \emph{control variables}. The calculation of the controls is governed by the minimization of a functional involving quantities of interest calculated in terms of the state and the control variables. This optimization problem is posed over a certain timeframe of interest, such that the terms in the functional can be calculated over the whole time period {via} (spatio)temporal integration to introduce the desired target dynamics. The functional terms can also be posed at the time horizon to define their terminal target value (i.e., an \emph{endpoint control}). In the first case, the integrand usually features a linear or a quadratic formulation. The linear case leads to controls that switch between their minimum and maximum admissible value (i.e., \emph{a bang-bang control}), whereas the quadratic case results in controls that smoothly adjust their value within their admissible range to the dynamics of the state variable (i.e., a \emph{continuous control}). While the admissible range of the controls determines their minimal and maximal value at each timepoint, the OCT problem may further feature an \emph{isoperimetric constraint} that limits the total value of a control variable over time.

\begin{figure*}[!t]
\centering
\includegraphics[width=\linewidth]{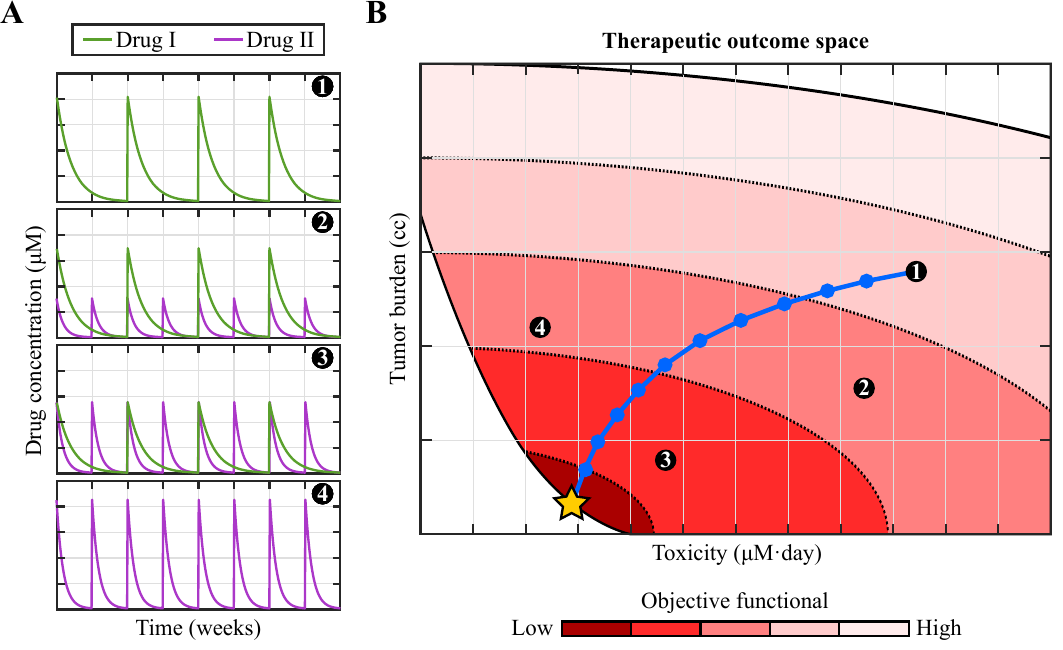}
\caption{\textbf{Optimal control theory enables the design of personalized cancer treatments.} Let us assume that we have a validated, personalized mechanistic model of tumor growth and treatment response. The standard-of-care treatment for this tumor consists of chemotherapy using drug I (delivered every two weeks), drug II (delivered weekly), or a combination of both drugs. (A) Temporal changes of the concentration in blood of either drug during the course of four illustrative standard-of-care chemotherapeutic regimens. (B) Space of therapeutic outcomes spanned by the tumor burden (e.g., the tumor volume) and the treatment toxicity (e.g., the total cumulative exposure to both drugs) calculated at a fixed time after the conclusion of chemotherapy by leveraging the personalized mechanistic model for any admissible combination of drugs I and II \cite{Jarrett2020,Colli2021,Wu2022}.
An objective functional combining the tumor burden and the toxicity further assesses the performance of each treatment regimen (color mapping in panel B), such that low values of this functional identify treatments achieving large tumor burden reductions while entailing low toxicity.
OCT usually involves an iterative algorithm whereby, starting from a standard-of-care regimen, each iteration adjusts the treatment towards an improved performance. The iterative strategy is represented by circles in the blue path, which starts in standard-of-care regimen 1 and ends in the optimal treatment for this scenario (golden star).
}
\label{fig_oct}
\end{figure*}

In the context of cancer, OCT has been employed to find optimal treatment regimens that maximize tumor control with minimal toxicities \cite{Lenhart2007, Lorenzo2023, Schattler2016}. Since these OCT applications have been mainly developed at tissue or organ scale, they usually rely on mechanistic models that are posed in terms of ODEs and
PDEs. The state variables usually correspond to the tumor volume \cite{Lima2022, Benzekry2013}, the total tumor cell count \cite{Lorenzo2023}, the tumor cell density \cite{Corwin2013}, the tumor phase field \cite{Colli2021}, and key substances related to tumor dynamics \cite{Colli2021, IzurzunArana2018}. The control variables can represent the treatments (e.g., the radiation or drug dosage delivered in each treatment session) \cite{Corwin2013,Lima2022, IzurzunArana2018} or their dose-dependent effects on tumor dynamics (e.g., therapeutically-induced tumor cell death) \cite{Colli2021, Bodzioch2021, Schattler2016, Benzekry2013}. Regarding the formulation of the functional, the endpoint terms usually aim at minimizing the tumor burden at the prescribed time horizon \cite{Colli2021, Bodzioch2021, Benzekry2013, Schattler2016}, although this goal can also be achieved dynamically over the OCT problem timeframe by leveraging (spatio)temporally-resolved integral terms \cite{Colli2021, Bodzioch2021, IzurzunArana2018, Lima2022}.
A clinically-admissible dosage range is usually defined to limit the amount of treatment delivered in every instant (e.g., the radiation or drug dose fraction) \cite{Corwin2013, Colli2021, Benzekry2013}, and the OCT functional usually incorporates additional (spatio)temporally-resolved integral terms that seek to minimize the
total dosage under the optimal regimen \cite{Colli2021, Lima2022, Schattler2016}. Additionally, isoperimetric constraints
can further bound the total amount of a therapeutic agent delivered over
the OCT problem timeframe \cite{Lima2022, Lenhart2007, Benzekry2013}.
Figure~\ref{fig_oct} provides an example of the OCT framework in the context of designing optimal, personalized chemotherapy regimens.

Several studies have shown that OCT is a promising framework to maximize treatment outcomes while minimizing the amount of therapeutic agents delivered to the patient, thereby minimizing potential toxicities and side-effects. For example, OCT has been leveraged to investigate optimal radiation plans for glioblastoma \cite{Corwin2013}, optimal combinations of cytotoxic and antiangiogenic therapies in solid tumors at the metastatic stage \cite{Benzekry2013} and advanced prostate cancer \cite{Colli2021}, optimal combinations of chemotherapy and HER2-targeted therapy for HER2+
breast cancer \cite{Lima2022}, optimal hormonal treatments for advanced prostate cancer \cite{IzurzunArana2018}, and optimal drug regimens for neodjuvant chemotherapy of locally advanced breast cancer \cite{Wu2022}. Nevertheless, a main limitation of OCT is that the resulting optimal control variables may correspond to therapeutic regimens that are not possible to achieve in the clinic. For example, the on-off treatment regimen resulting from a bang-bang control may
require maintaining a maximal concentration of a therapeutic for unfeasible times, while the smooth concentration dynamics obtained with a continuous control may not correspond to any known and clinically-validated therapeutic agent \cite{Colli2021, Bodzioch2021, Lorenzo2023}. Mathematical adjustments to the problem formulation may contribute to palliate these issues, such as defining the control as a timepoint therapeutic agent source within an ODE or PDE modeling its interaction with tumor dynamics and its natural decay \cite{Lima2022, IzurzunArana2018}. Alternatively, if the eligible treatments for an individual patient constitute a limited set of options, the OCT framework can be employed to evaluate each of them separately and choose the one rendering best results in terms of tumor control and total dosage of the therapeutic agent \cite{Wu2022}. Additionally, OCT can also be applied in a two-step strategy, whereby the resulting theoretical controls can serve as basis to define clinically feasible optimal therapies achieving comparable treatment outcomes as the parent OCT solution \cite{Colli2021, Bodzioch2021}. In the future, the development of new biochemical technologies enabling the control of local delivery and tissue concentration of drugs \cite{Chua2020, Desrosiers2022} as well as the production of therapeutic drugs with pharmacokinetics and pharmacodynamics tailored to each individual patient \cite{Iyengar2012, Shi2017} may open a new opportunity to apply OCT for the personalized design of treatment regimens. However, the clinical translation of OCT frameworks will still constitute a dramatically bigger challenge than that of patient-specific models for tumor forecasting because OCT would produce novel treatment regimens that may have not been assessed within a clinical study. Hence, the computationally-derived therapies obtained \emph{via} OCT would require a close monitoring to control treatment response, toxicity, and the well-being of every patient in the treatment arm, which can be extremely difficult to balance and maintain during a clinical study.

\subsection{Digital twins}

A digital twin can be defined as a virtual representation of a physical object (e.g., a human organ) by means of a computational model that is continuously informed by object-specific measurements to enable decision-making about the physical object based on its current and future status (e.g., health or response to external agents) \cite{Niederer2021, Wu2022a, Rasheed2020, Tao2019}. The initial setup of a digital twin usually requires data characterizing the properties of the constituents of the physical object (e.g., physiological features or geometry). Then, periodic or continuous measurements of the physical object status enable the
updating of the parameterization of the computational models at the core of the digital twin and, hence, assess the integrity and quality of performance of the physical object. Computational forecasts obtained with the updated models enable the projection of the behavior of the digital twin in the future, which can inform about potential adjustments to the physical object operation that are required to optimally maintain its health and satisfactory performance. 
Likewise, the response of the physical object to external actions 
can be optimized by further exploiting the computational forecasts of the underlying models (e.g., 
the delivery of a drug to control the tumor). Central to the construction of a digital twin is the selection of metrics that robustly and accurately characterize its status and performance. These metrics need to be readily calculated from the computational models by using the measurements from the physical object and serve in the decision-making process. 
Probabilistic digital twins cast these technologies in a Bayesian framework that enables to account for the uncertainty in the incoming data and model predictions \cite{Kapteyn2021, Wu2022a}, thereby allowing for a richer risk-based optimization of the physical object status and performance than digital twins based on deterministic approaches.

\begin{figure*}[!t]
\centering
\includegraphics[width=\linewidth]{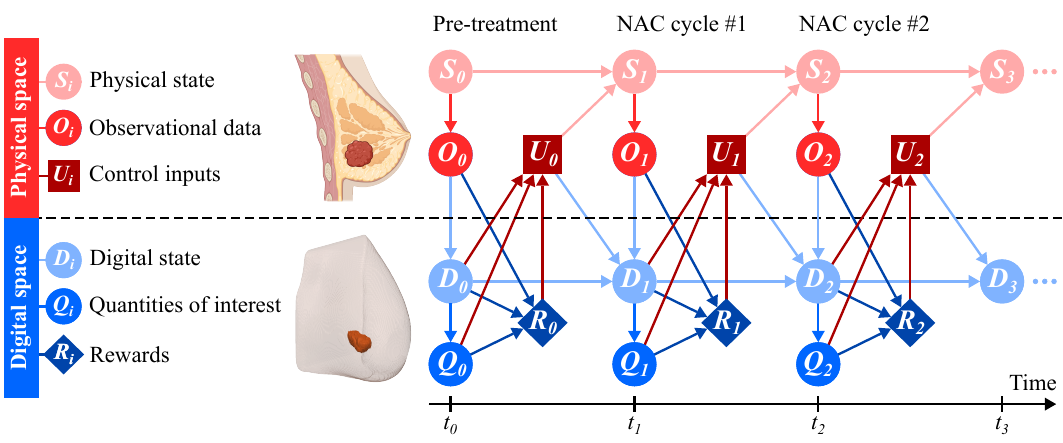}
\caption{\textbf{Digital twins provide a practical computational framework to optimally manage cancer monitoring and treatment on a patient-specific basis.}  
Personalized digital twin framework for neoadjuvant chemotherapy (NAC) of breast cancer \cite{Wu2022b,Wu2022a}. 
The \emph{physical twin} represents the patient's tumor and the host organ, while the \emph{digital twin} consists of a model that describes the spatiotemporal dynamics of tumor growth and treatment response.
The formulation of the digital twin requires the definition of three key quantities in the \emph{physical space} and another three key quantities in the \emph{digital space}. First, the \emph{physical state} ($S_i$) describes the physiology and local anatomy of the patient at time $t_i$.
The only information available on the physical state is provided by the observational data ($O_i$), which offer a limited and indirect assessment of the physical twin (e.g., \emph{via} imaging, biomarker, and omics data).
The calibration of the model constituting the digital twin to the observational data enables the determination of the digital state at time $t_i$ ($D_i$).
The calibrated model enables the calculation of quantities of interest ($Q_i$) for the design of each therapeutic cyle of the treatment (e.g., terminal total tumor cell count or volume, toxicity or time to progression).
The rewards ($R_i$) quantify the performance of a candidate therapeutic regimen on the basis of the current observational data, digital state, and the corresponding quantities of interest. 
The formulation of the rewards usually enables the definition of an optimization problem to find the best treatment plan for each patient, which would then be delivered to the patient as a control input ($U_i$).
Finally, the digital twin can be progressively updated as new data from the patient becomes available, which would enable the personalized adjustment of the treatment plan to optimize therapeutic outcomes. The image for the physical twin was created using BioRender, and the image of the digital twin is courtesy of Dr. Chengyue Wu.
}
\label{fig_dtwins}
\end{figure*}

In the context of medicine, digital twins have been developed to address several procedures \cite{Niederer2021, Wu2022a}, such as the planning of surgeries and interventions \cite{Malone2010, Brunet2019, Niederer2021, CorralAcero2020}, the convection-enhanced delivery of drugs to treat brain tumors and neurological disorders \cite{Woodall2021, WembacherSchroeder2021}, and the control of blood levels of glucose in diabetic patients \cite{Shamanna2020, Lal2019}. The success of
biomechanistic models to recapitulate and predict the growth and treatment response of different types of tumors in individual patients motivates the development of digital twins for applications in clinical oncology \cite{Wu2022a, HernandezBoussard2021}. In particular, digital
twins have been proposed as a promising computational environment to effectively translate these models to the clinic and guide biological discovery in the preclinical setting. For instance, preliminary efforts have shown the application of digital twins in monitoring and predicting the response of breast cancer to neoadjuvant therapy in individual patients \cite{Wu2022b},  experimental planning to investigate the mechanical cues underlying tumor growth \emph{in vitro} \cite{Urcun2021}, and investigating the combination of chemotherapy and antiangiogenic therapy in xenograft animal models \cite{Hadjicharalambous2022}. As an example, Fig.~\ref{fig_dtwins} provides an overview of a prototype of a digital twin to optimally monitor and design neaodjuvant chemohterapy for breast cancer. In the future, increased access to data and computational resources may contribute to the development of these pioneering digital twin prototypes towards a technology that can be assessed in a clinical trial for its ultimate deployment as a predictive, personalized decision-making technology to guide the oncological management of tumors \cite{Wu2022a, HernandezBoussard2021}.



\section{MECHANISTIC MODELS INCORPORATING AI}

Given the limitations of stand-alone AI  and mechanistic models, it is essential to consider the potential of combining both approaches. This opens up new avenues for improved understanding of tumor growth and development of personalized treatment strategies \cite{Mascheroni2021,benzekry2020artificial,Hormuth2021}. Below we describe a few possibilities to incorporate AI into mechanistic models.

\subsection{Parameter identification and initialization}

One key drawback of AI and machine learning approaches designed to assess tumor growth and dynamics is overfitting to the training set, which can be largely attributed to a combination of sparse clinical data that is limited by a patient’s presentation, and suboptimal methodological approaches. Solid tumors exhibit substantial inter- and intra-tumor heterogeneity, and AI models frequently fail to capture the full complexity and extent of this heterogeneity. This can be addressed by generating patient-specific, mechanistic models of tumor dynamics which are initialized and parameterized by data obtained using deep neural networks trained on a larger training set.
\begin{figure*}
\centering
\includegraphics[width=\linewidth]{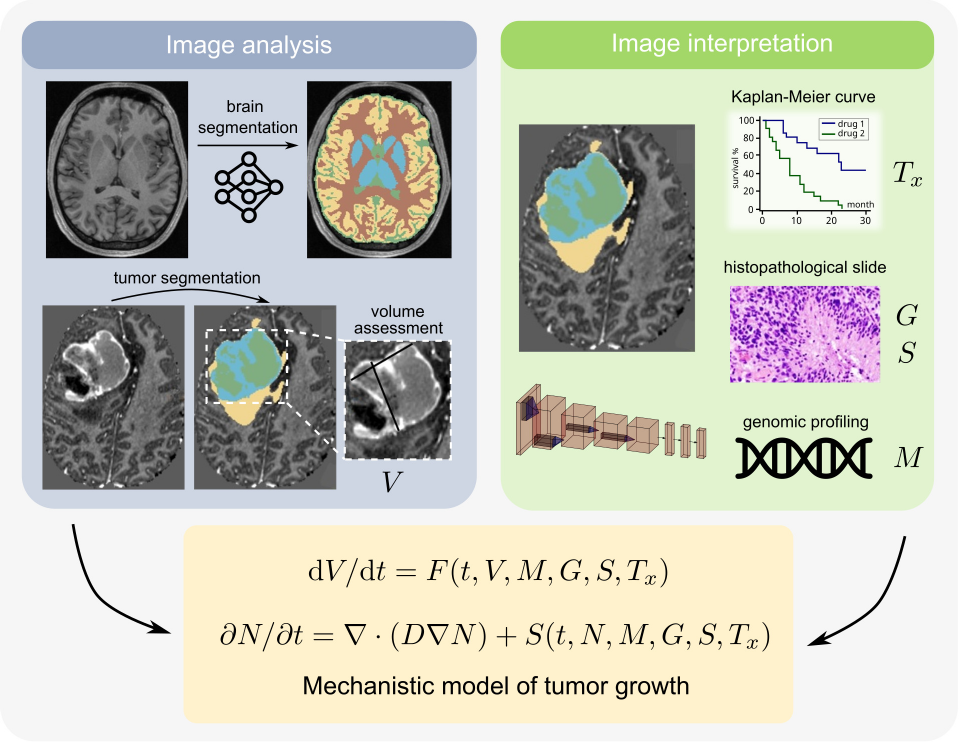}
\caption{\textbf{Initializing mechanistic models with parameters from AI models.} Potential approach to combine mechanistic models with AI models in a clinical scenario relevant to brain cancer. We propose to utilize trained glioma segmentation models \cite{patel2021opportunities} for volumetric quantification of tumors [left], while classification networks can be used for oncogenic driver/molecular marker status prediction (M), histopathological grading and staging of the neoplasm (G, S), and assess treatment response (Tx) [right]. Each of these can serve as parameters and inputs to the patient-specific mechanistic model, which can take the form of and ODE or a PDE.}
\label{fig_param_init}
\end{figure*}
These parameters could be the volumetric quantification outputs of deep learning segmentation models, as well as classification models designed to identify other contributors to tumor growth and regression, which may include but are not limited to oncogenic driver/molecular marker status of tumors, any therapeutic treatment administered and the corresponding dose response, as well as the grading and staging of disease; see Fig.~\ref{fig_param_init}. A wide range of input data can be utilized to identify these parameters, such as genetic profiles, imaging results and clinical history. The parameters can be fed into a mechanistic model based on ODEs \cite{Benzekry2014, Yin2019} or PDEs \cite{Wu2022b, Hormuth2021, Lorenzo2023}. Upon training on a sufficiently large dataset, these segmentation or classification networks can be used to run inference on individual patients, and generate initial patient-specific parameters for the mechanistic models. The mechanistic models can subsequently be run to forecast tumor growth at various time points, as well as assess treatment response and shed light on other diagnostically and prognostically meaningful variables. This integration allows for more personalized tumor growth predictions and treatment planning that accounts for inter-patient variability.

\subsection{Surrogate modeling}

Surrogate models serve as approximations of complex mechanistic models and are particularly useful (1) when fast solving is required due to minimal time, and (2) when many simulations are required, such as in the case of conducting uncertainty analyses \cite{sobester2008engineering}. AI and deep learning can be utilized to develop surrogate models that approximate the behavior of mechanistic models. These surrogate models can be trained on the outputs of mechanistic models, which enables the AI model to learn the mapping between the input parameters of the mechanistic model and the corresponding mechanistic model predictions. The recent explosive growth of neuro-symbolic AI \cite{garcez2023neurosymbolic} is particularly relevant in the context of surrogate modelling: architectures such as transformers \cite{khan2022transformers,tay2022efficient} offer both predictive accuracy and computational efficiency that are crucial elements to creating a surrogate model. Once trained, surrogate models can provide tumor growth predictions or conduct extensive simulations with prognostic value.

\begin{boxH}
GROWTH OF NEURO-SYMBOLIC AI\\
Symbolic machine learning involves the manipulation of symbols during a discrete search process to find the optimal representation for solving a specific classification or regression task. By combining neural and symbolic AI architectures, neuro-symbolic AI aims to tackle the inherent strengths and weaknesses of both approaches. This integration results in a robust artificial intelligence system that possesses the ability to reason, learn, and perform cognitive modeling tasks effectively. Large language models (LLM) such as GPT-3 utilize neuro-symbolic AI, which uses transformers as their architecture. \\[.2cm]
Transformers:\\
Transformers employ a self-attention mechanism to capture contextual relationships between words or tokens, enabling them to generate high-quality language representations. They have become the backbone of numerous Natural Language Processing (NLP) applications, computer vision and multi-modal processing, and have been prevalently adopted as the state-of-the-art architecture for training LLMs.
\end{boxH}


In the first case, a critical situation where fast solving is essential is in surgical procedures, where quick tumor growth predictions are required to inform surgical decision making. Surrogate models can quickly generate predictions of tumor response to different therapeutic interventions, enabling real-time decision-making. These predictions can subsequently be used by surgeons to guide their specific surgical approach and optimize patient outcomes.

Uncertainty analysis, in the oncology space, involves quantifying the impact of uncertain parameters on tumor dynamics. In situations where uncertainty analysis or sensitivity analysis requires running multiple simulations, AI models can be used to construct surrogate models that can efficiently handle the computational load. Conceptually, this is analogous to the creation of a digital twin \cite{Wu2022a} with the notable difference that the computations are performed by a neural network. AI models can also enhance the efficiency of uncertainty quantification methods such as Monte Carlo simulations or optimization algorithms.

The effectiveness of surrogate models relies on the availability of high-quality data and accurate mechanistic models. Sufficient training data, covering diverse patient cases, is necessary to ensure that surrogate models capture the heterogeneity of tumor growth. The underlying mechanistic models also need to be carefully validated and refined to accurately represent the corresponding biological processes. Ultimately, the integration of AI and mechanistic models requires individuals with expertise in both domains to come together and effectively optimize their interactions.

\subsection{Biology-informed neural networks}

Recent research has shown that AI algorithms have achieved a noteworthy level of success predicting patient outcome and other prognostic endpoints using sequences of longitudinal images for a single patient \cite{xu2019deep}. However, longitudinal images for a single patient are rarely available in standard-of-care cancer protocols, and it is unlikely that they will become available in the near future. One way to alleviate this challenge is combining these purely data-driven approaches with a biomechanistic model. Physics-informed neural networks (PINNs) \cite{raissi2019physics}, and their extension to medical sciences, namely biology-informed neural networks (BINNs),  \cite{kuenzi2020predicting,lagergren2020biologically} constitute an ideal mathematical framework to accomplish this combination. The recent emergence of recurrent neural networks in medical sciences \cite{xu2019deep} offers an opportunity for the development of BINNs that couple machine learning with evolutive mechanistic models based on PDEs. For example, the recurrent neural network illustrated in Fig.~\ref{fig_lung} could be enhanced by adding to the loss function a term that penalizes deviations from solutions to a biomechanistic PDE with uncertain parameters. We believe this is a very promising area of future research. The framework of BINNs can also be exploited to the discovery of hidden biomechanistic PDEs following an approach similar to that illustrated in \cite{raissi2019physics}.

\section{FIVE FUNDAMENTAL CHALLENGES}

\subsection{Predicting metastasis}

Despite major advances in cancer diagnosis and treatment, the vast majority of cancer deaths are the result of metastatic disease \cite{RN7,RN8,RN9}. During metastasis, tumor cells interact with multiple distinct microenvironments to facilitate the process of leaving the primary tumor site and forming discontinuous secondary masses. Due to the high morbidity rate, there has been a growing emphasis on developing model systems that specifically study metastasis. However, many of these model systems equate the process of cell motility and invasion to metastasis. This assumption is incorrect \cite{RN10}. Recent studies have demonstrated that cells with high invasive capacity typically are only capable of seeding distal tissues and are not capable of completing the metastatic cascade \cite{RN10,RN11,RN12}. It is becoming increasingly clear that models must account for cellular plasticity to predict metastasis. To help bring clarity to this issue 4 key hallmarks of metastasis have recently been defined and include \cite{RN13}: (i) the ability to modulate both the secondary premetastatic site and the local primary tumor microenvironment, (ii) cellular plasticity, (iii) motility and invasion, and (iv) the ability to colonize new tissues \cite{RN13}. The need to understand the dynamics at both the primary site and metastatic site creates a data challenge in that metastatic tissues are not commonly collected from patients, which limits the ability to use omic-based approaches to study the dynamics of the tissue. Furthermore, metastasis is a rare event, which limits the pool of cell lines and animal model systems available for data collection. Recent advances in microphysiological systems \cite{RN14,RN15,RN16} and \emph{ex vivo} cultures \cite{RN17} provide a promising route in which the dynamic events that occur during metastasis can be studied under controlled setting to inform \emph{in silico} models. These platforms allow for systematic evaluation of cellular heterogeneity \cite{RN12,RN18,RN19}, tissue histological factors that affect migration \cite{RN20} as well as mechanical forces within metastatic tissues and their effect on cell seeding and colonization \cite{RN16}. 

\subsection{Reliability of the predictions}

The last few decades have witnessed an increase in the sophistication of computational tumor models which has been driven by the growth of our scientific understanding of tumor biology and computer power. Transferring highly-complex models to the clinic requires very high levels of model reliability and robustness. To ensure reliability in safety-critical applications like treatment planning, uncertainties in parameters, initial and boundary conditions, numerical approximations as well as the natural stochasticity of the system must be quantified. Uncertainty quantification techniques, including sensitivity analysis, parameter estimation, Bayesian inference, probabilistic modeling, and Monte Carlo simulations, help assess reliability, explore outcomes, and estimate confidence intervals \cite{BradyNicholls2020, Zahid2021}. In tumor growth modeling, uncertainty quantification aids in understanding model limitations and uncertainties, improving decision-making in treatment planning, and guiding prognosis assessment. It also enhances model robustness and generalizability, driving refinement and improvement of modeling approaches. 

\subsection{Practicality of the predictions}

There are several key barriers on both the computational and clinical fronts that limit practical or clinically-actionable predictions. On the computational front, predictions based on medical imaging currently require multiple manual or semi-automated steps to prepare patient data for modeling.  These steps include tumor and tissue segmentation, verification of image registration, definition of the computational domain, and incorporating clinical notes and treatment schedules. Although automated approaches could potentially be used for many of these tasks, their ability to handle variations in data quality, data type, and treatment effects need to be demonstrated. An additional computational challenge is calibration and personalization of mathematical models of tumor growth in a timely fashion. Depending on the number of parameters, the size of the computational domain, and the calibration approach, models could take greater than a day for a single calibration \cite{jarrett_quantitative_2021}. Reduced order modeling approaches have the potential to address this barrier \cite{doi:10.1137/130932715}.

On the clinical front, it is crucial for model predictions of optimal therapeutic regimens be made available within a clinically-actionable time frame. These predictions should also recommend regimens that are reasonable for patients to complete and can be readily integrated into clinical workflows. Moreover, while confidence in the use of mathematical modeling in the guiding of patient care is growing, it is important for models to recommend optimal regimens that do not deviate greatly outside of what clinicians would typically select or are comfortable for delivering. Thus, close collaboration with clinicians from the earliest stages of a given study is absolutely required. Radiotherapy provides guidance on how to proceed as there are many approaches for personalizing dose maps based on imaging data  \cite{jaffray2012image}; for example, potential treatment regimens could be limited to dose escalation or de-escalation simply within the tumor region of interest rather than deviation from the standard-of-care treatment schedule.

\subsection{Incorporating multiscale data into models}

We contend that in the next decade, tumor modeling will continue to be driven by data availability. Arguably, the most successful models will be those with greatest ability to integrate more data types consistently in a reliable and systematic manner. Genomic, proteomic, microscopy and imaging data have different nature and come at different length scales which poses a formidable task to modelers. Models that incorporate all these data types will necessarily be complex and difficult to interpret. The lack of interpretability is inherent to multiscale models with emerging behavior  that arises from the interactions between components at different scales. These emergent properties may not be directly observable or intuitive at individual scales, making it difficult to identify and transfer the underlying mechanisms that give rise to them. Therefore tumor models incorporating multiscale data will require extensive validation, verification and uncertainty quantification. Also, the standard approach to determine model parameters in engineering and physics has been to design a specific experiment that would reveal the value of a single parameter. This strategy is likely to be unsuccessful in tumor models that incorporate multiscale data. However, given the current progress in inverse problems fueled by machine learning, we envision algorithms capable of identifying all parameters at once, given large amounts of data. Another critical challenge to be overcome is finding an adequate scale separation and a systematic way to transfer data and model outputs across scales.

\subsection{Consistency and bias}
A desirable characteristic of any predictive model is consistency of predictions. Consistency, or repeatability refers to the ability of a model to generate near-identical predictions for the same patient under identical conditions, ensuring that the model produces precise, reliable outputs in the clinical setting. AI models in the existing literature frequently suffer from a lack of reliability due to overfitting: multiple images obtained from the same patient under similar or identical settings tend to result in different model outcomes. A number of approaches have been utilized to address this concern, such as dedicated optimization of models for repeatability \cite{ahmed2022reproducible}, loss function optimization \cite{ahmed2022focal}, and Monte Carlo (MC) dropout \cite{lemay2022improving}. Linking AI model predictions with patient-specific, mechanistic models can make modeling of tumor dynamics more patient specific and thus more consistent within each patient.

\begin{boxH}
IMPROVING REPEATABILITY OF DEEP LEARNING MODELS\\
Recent work has shed light onto the importance of consistency and repeatability of deep learning model predictions as a crucial and optimizable task \cite{ahmed2022reproducible,ahmed2022focal,lemay2022improving}. Model hyperparameters such as the loss function and the model architecture can be chosen based on model performance on key repeatability metrics such as weighted kappa \cite{cohen1960coefficient}. In particular, recent work has shown that focal loss, which is a modulated form of cross-entropy loss, and Monte Carlo dropout \cite{ahmed2022focal,lemay2022improving}, which is a model regularization technique involving training and testing a neural network with dropout, can improve repeatability of model predictions.
\end{boxH}


Another key concern with AI and ``big data" is the issue of bias. Frequently, training data is not representative of patient-level heterogeneity or variability that is encountered with real-world clinical data. Recent work has reported systemic biases in AI model predictions and lead to categorically biased recommendations \cite{gichoya2022ai,obermeyer2019dissecting}. Given that AI model recommendations are intended to be a part of clinical-decision making flow-diagrams, an incorrect, unreliable or biased model prediction would lead to a particular cascade of undesirable downstream clinical actions, that might significantly jeopardize the health and safety of a patient, and put their lives at risk. Combining patient-level mechanistic models with AI model predictions addresses the issue of bias to some extent by accounting for the patient-level heterogeneity. However, the AI models that are used for parameter initialization or as surrogate models can still output biased predictions, if not trained on a sufficiently large and multiply heteregoneous dataset.

\begin{boxA}
SUMMARY POINTS\\
\begin{enumerate}
\item Mechanistic or biology-based models of tumor growth and response provide ``interpretable" model predictions whose model parameters are rooted in the key components of tumor biology. 
\item AI has shown success in cancer diagnosis and treatment response assessment, but models are complex and lack interpretability.
\item Quantitative medical imaging techniques are clinically feasible and report on underlying tumor biology. 
\item Optimal control theory can be applied to enable the design of personalized cancer treatments that balances multiple constraints. 
\item Digital twin approaches can be to applied optimally manage cancer monitoring and treatment on a patient-specific basis. 
\item AI approaches, such as deep learning segmentation and radiogenomic classification, can be used to identify and initialize input parameters for mechanistic models.
\item AI models can serve as surrogate models that approximate the behavior of mechanistic models, particularly when fast solving is required and when many simulations need to be performed (e.g., for uncertainty analysis).
\item While the combination of AI with mechanistic models represents a promising advance, limitations such as metastasis prediction, multiscale data incorporation, lack of consistency, and bias need to be considered. 
\end{enumerate}
\end{boxA}


\section*{DISCLOSURE STATEMENT}
The authors are not aware of any affiliations, memberships, funding, or financial holdings that might be perceived as affecting the objectivity of this review. 

\section*{AUTHORS CONTRIBUTIONS}
All authors were responsible for development of the research and concepts presented; wrote sections of the paper; provided critical review, insights, and revisions to the presented; and read and approved the final paper.

\section*{ACKNOWLEDGMENTS}
G.L. acknowledges funding from the European Union’s Horizon 2020 research and innovation programme under the Marie Sk\l{}odowska-Curie grant agreement No. 838786. D.A.H. acknowledges funding from Cancer Prevention and Research Institute of Texas RP 220225. T.E.Y. acknowledges funding from the National Cancer Institute for funding through 1U24 CA226110, 1R01 CA235800, 1U01CA253540, 1R01CA260003, and 1R01CA276540. T.E.Y. acknowledges funding from the Cancer Prevention and Research Institute of Texas \emph{via} CPRIT RR160005 and he is a CPRIT Scholar of Cancer Research. H.G. was partially supported by the National Science Foundation through awards 1852285 and 1952912. The opinions, findings, and conclusions, or recommendations expressed are those of the author(s) and do not necessarily reflect the views of the National Science Foundation. H.G. was partially supported by the Purdue Center for Cancer Research. 

\bibliographystyle{plain}
\bibliography{references.bib}

\begin{thebibliography}{100}

\bibitem{ahmed2022reproducible}
Syed~Rakin Ahmed, Brian Befano, Andreanne Lemay, Didem Egemen, Ana~Cecilia
  Rodriguez, Sandeep Angara, Kanan Desai, Jose Jeronimo, Sameer Antani, Nicole
  Campos, et~al.
\newblock Reproducible and clinically translatable deep neural networks for
  cervical screening.
\newblock {\em medRxiv}, pages 2022--12, 2022.

\bibitem{ahmed2022focal}
Syed~Rakin Ahmed, Andreanne Lemay, Katharina~V Hoebel, and Jayashree
  Kalpathy-Cramer.
\newblock Focal loss improves repeatability of deep learning models.
\newblock In {\em Medical Imaging with Deep Learning}, 2022.

\bibitem{Alber2019}
Mark Alber, Adrian Buganza~Tepole, William~R Cannon, Suvranu De, Salvador
  Dura-Bernal, Krishna Garikipati, George Karniadakis, William~W Lytton, Paris
  Perdikaris, Linda Petzold, et~al.
\newblock Integrating machine learning and multiscale modeling—perspectives,
  challenges, and opportunities in the biological, biomedical, and behavioral
  sciences.
\newblock {\em NPJ digital medicine}, 2(1):115, 2019.

\bibitem{alcorn2019strike}
Michael~A Alcorn, Qi~Li, Zhitao Gong, Chengfei Wang, Long Mai, Wei-Shinn Ku,
  and Anh Nguyen.
\newblock Strike (with) a pose: Neural networks are easily fooled by strange
  poses of familiar objects.
\newblock In {\em Proceedings of the IEEE/CVF conference on computer vision and
  pattern recognition}, pages 4845--4854, 2019.

\bibitem{anderson_mathematical_2018}
Alexander R.~A. Anderson and Philip~K. Maini.
\newblock Mathematical oncology.
\newblock {\em Bulletin of Mathematical Biology}, 80(5):945--953, 2018.

\bibitem{Angeli2018}
Stelios Angeli, Kyrre~E. Emblem, Paulina Due-Tonnessen, and Triantafyllos
  Stylianopoulos.
\newblock Towards patient-specific modeling of brain tumor growth and formation
  of secondary nodes guided by {DTI-MRI}.
\newblock {\em NeuroImage: Clinical}, 20:664--673, 2018.

\bibitem{RN8}
S.~Badve, D.~J. Dabbs, S.~J. Schnitt, F.~L. Baehner, T.~Decker, V.~Eusebi,
  S.~B. Fox, S.~Ichihara, J.~Jacquemier, S.~R. Lakhani, J.~Palacios, E.~A.
  Rakha, A.~L. Richardson, F.~C. Schmitt, P.~H. Tan, G.~M. Tse, B.~Weigelt,
  I.~O. Ellis, and J.~S. Reis-Filho.
\newblock Basal-like and triple-negative breast cancers: a critical review with
  an emphasis on the implications for pathologists and oncologists.
\newblock {\em Mod Pathol}, 24(2):157--67, 2011.

\bibitem{baid2021rsna}
Ujjwal Baid, Satyam Ghodasara, Suyash Mohan, Michel Bilello, Evan Calabrese,
  Errol Colak, Keyvan Farahani, Jayashree Kalpathy-Cramer, Felipe~C Kitamura,
  Sarthak Pati, et~al.
\newblock The rsna-asnr-miccai brats 2021 benchmark on brain tumor segmentation
  and radiogenomic classification.
\newblock {\em arXiv preprint arXiv:2107.02314}, 2021.

\bibitem{Baid2020}
Ujjwal Baid, Swapnil~U. Rane, Sanjay Talbar, Sudeep Gupta, Meenakshi~H. Thakur,
  Aliasgar Moiyadi, and Abhishek Mahajan.
\newblock {Overall Survival Prediction in Glioblastoma With Radiomic Features
  Using Machine Learning}.
\newblock {\em Frontiers in Computational Neuroscience}, 14:61, aug 2020.

\bibitem{bakas2017segmentation2}
S~Bakas, H~Akbari, A~Sotiras, M~Bilello, M~Rozycki, J~Kirby, J~Freymann,
  K~Farahani, and C~Davatzikos.
\newblock Segmentation labels for the pre-operative scans of the tcga-gbm
  collection (2017), 2017.

\bibitem{bakas2017segmentation}
S~Bakas, H~Akbari, A~Sotiras, et~al.
\newblock Segmentation labels for the pre-operative scans of the tcga-gbm
  collection. the cancer imaging archive, 2017.

\bibitem{bakas2017advancing}
Spyridon Bakas, Hamed Akbari, Aristeidis Sotiras, Michel Bilello, Martin
  Rozycki, Justin~S Kirby, John~B Freymann, Keyvan Farahani, and Christos
  Davatzikos.
\newblock Advancing the cancer genome atlas glioma mri collections with expert
  segmentation labels and radiomic features.
\newblock {\em Scientific data}, 4(1):1--13, 2017.

\bibitem{baker_workshop_2019}
Nathan Baker, Frank Alexander, Timo Bremer, Aric Hagberg, Yannis Kevrekidis,
  Habib Najm, Manish Parashar, Abani Patra, James Sethian, Stefan Wild, et~al.
\newblock Workshop report on basic research needs for scientific machine
  learning: Core technologies for artificial intelligence.
\newblock Technical report, USDOE Office of Science (SC), Washington, DC
  (United States), 2019.

\bibitem{Baldock2013}
Anne Baldock, Russell Rockne, Addie Boone, Maxwell Neal, Carly Bridge, Laura
  Guyman, Maceij Mrugala, Jason Rockhill, Kristin~Rae Swanson, Andrew~Daniel
  Trister, A~Hawkins-Daarud, and David~M Corwin.
\newblock From patient-specific mathematical neuro-oncology to precision
  medicine.
\newblock {\em Frontiers in Oncology}, 3, 2013.
\newblock {ISBN}: 2234-943X.

\bibitem{bansal2020sam}
Naman Bansal, Chirag Agarwal, and Anh Nguyen.
\newblock Sam: The sensitivity of attribution methods to hyperparameters.
\newblock In {\em Proceedings of the ieee/cvf conference on computer vision and
  pattern recognition}, pages 8673--8683, 2020.

\bibitem{baxi2022digital}
Vipul Baxi, Robin Edwards, Michael Montalto, and Saurabh Saha.
\newblock Digital pathology and artificial intelligence in translational
  medicine and clinical practice.
\newblock {\em Modern Pathology}, 35(1):23--32, 2022.

\bibitem{doi:10.1137/130932715}
Peter Benner, Serkan Gugercin, and Karen Willcox.
\newblock A survey of projection-based model reduction methods for parametric
  dynamical systems.
\newblock {\em SIAM Review}, 57(4):483--531, 2015.

\bibitem{benzekry2020artificial}
S{\'e}bastien Benzekry.
\newblock Artificial intelligence and mechanistic modeling for clinical
  decision making in oncology.
\newblock {\em Clinical Pharmacology \& Therapeutics}, 108(3):471--486, 2020.

\bibitem{Benzekry2013}
S\'{e}bastien Benzekry and Philip Hahnfeldt.
\newblock Maximum tolerated dose versus metronomic scheduling in the treatment
  of metastatic cancers.
\newblock {\em Journal of Theoretical Biology}, 335:235--244, 2013.

\bibitem{Benzekry2014}
S\'{e}bastien Benzekry, Clare Lamont, Afshin Beheshti, Amanda Tracz, John M.~L.
  Ebos, Lynn Hlatky, and Philip Hahnfeldt.
\newblock Classical mathematical models for description and prediction of
  experimental tumor growth.
\newblock {\em PLOS Computational Biology}, 10(8):e1003800, 2014.

\bibitem{blanco2023mechanotransduction}
B~Blanco, H~Gomez, J~Melchor, R~Palma, J~Soler, and G~Rus.
\newblock Mechanotransduction in tumor dynamics modeling.
\newblock {\em Physics of Life Reviews}, 2023.

\bibitem{Bodzioch2021}
Mariusz Bodzioch, Piotr Bajger, and Urszula Fory{\'s}.
\newblock Angiogenesis and chemotherapy resistance: optimizing chemotherapy
  scheduling using mathematical modeling.
\newblock {\em Journal of Cancer Research and Clinical Oncology},
  147(8):2281--2299, 2021.

\bibitem{box1979robustness}
George~EP Box.
\newblock Robustness in the strategy of scientific model building.
\newblock In {\em Robustness in statistics}, pages 201--236. Elsevier, 1979.

\bibitem{BradyNicholls2020}
Renee Brady-Nicholls, John~D Nagy, Travis~A Gerke, Tian Zhang, Andrew~Z Wang,
  Jingsong Zhang, Robert~A Gatenby, and Heiko Enderling.
\newblock Prostate-specific antigen dynamics predict individual responses to
  intermittent androgen deprivation.
\newblock {\em Nature Communications}, 11(1):1750, 2020.

\bibitem{Brummer2023}
Alexander~B. Brummer, Agata Xella, Ryan Woodall, Vikram Adhikarla, Heyrim Cho,
  Margarita Gutova, Christine~E. Brown, and Russell~C. Rockne.
\newblock Data driven model discovery and interpretation for {CAR T-cell}
  killing using sparse identification and latent variables.
\newblock {\em Frontiers in Immunology}, 14, 2023.

\bibitem{Brunet2019}
Jean-Nicolas Brunet, Andrea Mendizabal, Antoine Petit, Nicolas Golse, Eric
  Vibert, and St{\'e}phane Cotin.
\newblock Physics-based deep neural network for augmented reality during liver
  surgery.
\newblock In {\em Medical Image Computing and Computer Assisted
  Intervention--MICCAI 2019: 22nd International Conference, Shenzhen, China,
  October 13--17, 2019}, pages 137--145. Springer, Cham, 2019.

\bibitem{bull_hallmarks_2022}
Joshua~Adam Bull and Helen~Mary Byrne.
\newblock The hallmarks of mathematical oncology.
\newblock {\em Proceedings of the {IEEE}}, pages 1--18, 2022.

\bibitem{castell_quantitative_2008}
F~Castell and G~J~R Cook.
\newblock Quantitative techniques in 18fdg {PET} scanning in oncology.
\newblock {\em British Journal of Cancer}, 98(10):1597--601, 2008.

\bibitem{chang2018residual}
Ken Chang, Harrison~X Bai, Hao Zhou, Chang Su, Wenya~Linda Bi, Ena Agbodza,
  Vasileios~K Kavouridis, Joeky~T Senders, Alessandro Boaro, Andrew Beers,
  et~al.
\newblock Residual convolutional neural network for the determination of idh
  status in low-and high-grade gliomas from mr imagingneural network for
  determination of idh status in gliomas.
\newblock {\em Clinical Cancer Research}, 24(5):1073--1081, 2018.

\bibitem{10.1093/neuonc/noz106}
Ken Chang, Andrew~L Beers, Harrison~X Bai, James~M Brown, K~Ina Ly, Xuejun Li,
  Joeky~T Senders, Vasileios~K Kavouridis, Alessandro Boaro, Chang Su,
  Wenya~Linda Bi, and et~al.
\newblock {Automatic assessment of glioma burden: a deep learning algorithm for
  fully automated volumetric and bidimensional measurement}.
\newblock {\em Neuro-Oncology}, 21(11):1412--1422, 2019.

\bibitem{Chua2020}
Corrine Ying~Xuan Chua, Jeremy Ho, Sandra Demaria, Mauro Ferrari, and
  Alessandro Grattoni.
\newblock Emerging technologies for local cancer treatment.
\newblock {\em Advanced Therapeutics}, 3(9):2000027, 2020.

\bibitem{cohen1960coefficient}
Jacob Cohen.
\newblock A coefficient of agreement for nominal scales.
\newblock {\em Educational and psychological measurement}, 20(1):37--46, 1960.

\bibitem{Colli2021}
Pierluigi Colli, Hector Gomez, Guillermo Lorenzo, Gabriela Marinoschi,
  Alessandro Reali, and Elisabetta Rocca.
\newblock Optimal control of cytotoxic and antiangiogenic therapies on prostate
  cancer growth.
\newblock {\em Mathematical Models and Methods in Applied Sciences},
  31(07):1419--1468, 2021.

\bibitem{CorralAcero2020}
Jorge Corral-Acero, Francesca Margara, Maciej Marciniak, Cristobal Rodero,
  Filip Loncaric, Yingjing Feng, Andrew Gilbert, Joao~F Fernandes, Hassaan~A
  Bukhari, Ali Wajdan, Manuel~Villegas Martinez, Mariana~Sousa Santos, Mehrdad
  Shamohammdi, Hongxing Luo, Philip Westphal, Paul Leeson, Paolo DiAchille,
  Viatcheslav Gurev, Manuel Mayr, Liesbet Geris, Pras Pathmanathan, Tina
  Morrison, Richard Cornelussen, Frits Prinzen, Tammo Delhaas, Ada Doltra,
  Marta Sitges, Edward~J Vigmond, Ernesto Zacur, Vicente Grau, Blanca
  Rodriguez, Espen~W Remme, Steven Niederer, Peter Mortier, Kristin McLeod,
  Mark Potse, Esther Pueyo, Alfonso Bueno-Orovio, and Pablo Lamata.
\newblock {The {‘Digital Twin’} to enable the vision of precision
  cardiology}.
\newblock {\em European Heart Journal}, 41(48):4556--4564, 03 2020.

\bibitem{Corwin2013}
David Corwin, Clay Holdsworth, Russell~C. Rockne, Andrew~D. Trister, Maciej~M.
  Mrugala, Jason~K. Rockhill, Robert~D. Stewart, Mark Phillips, and Kristin~R.
  Swanson.
\newblock Toward patient-specific, biologically optimized radiation therapy
  plans for the treatment of glioblastoma.
\newblock {\em PLOS ONE}, 8(11):e79115, 2013.

\bibitem{coudray2018classification}
Nicolas Coudray, Paolo~Santiago Ocampo, Theodore Sakellaropoulos, Navneet
  Narula, Matija Snuderl, David Feny{\"o}, Andre~L Moreira, Narges Razavian,
  and Aristotelis Tsirigos.
\newblock Classification and mutation prediction from non--small cell lung
  cancer histopathology images using deep learning.
\newblock {\em Nature medicine}, 24(10):1559--1567, 2018.

\bibitem{Deisboeck2011}
Thomas~S. Deisboeck, Zhihui Wang, Paul Macklin, and Vittorio Cristini.
\newblock Multiscale cancer modeling.
\newblock {\em Annual Review of Biomedical Engineering}, 13(1):127--155, 2011.

\bibitem{Desrosiers2022}
Arnaud Desrosiers, Rabeb~Mouna Derbali, Sami Hassine, J{\'e}r{\'e}mie Berdugo,
  Val{\'e}rie Long, Dominic Lauzon, Vincent De~Guire, C{\'e}line Fiset, Luc
  DesGroseillers, Jeanne Leblond~Chain, et~al.
\newblock Programmable self-regulated molecular buffers for precise sustained
  drug delivery.
\newblock {\em Nature Communications}, 13(1):6504, 2022.

\bibitem{donahue2015long}
Jeffrey Donahue, Lisa Anne~Hendricks, Sergio Guadarrama, Marcus Rohrbach,
  Subhashini Venugopalan, Kate Saenko, and Trevor Darrell.
\newblock Long-term recurrent convolutional networks for visual recognition and
  description.
\newblock In {\em Proceedings of the IEEE conference on computer vision and
  pattern recognition}, pages 2625--2634, 2015.

\bibitem{eisenhauer_new_2009}
E~A Eisenhauer, P~Therasse, J~Bogaerts, L~H Schwartz, D~Sargent, R~Ford,
  J~Dancey, S~Arbuck, S~Gwyther, M~Mooney, L~Rubinstein, L~Shankar, L~Dodd,
  R~Kaplan, D~Lacombe, and J~Verweij.
\newblock New response evaluation criteria in solid tumours: revised {RECIST}
  guideline (version 1.1).
\newblock {\em European journal of cancer (Oxford, England : 1990)},
  45(2):228--247, 2009.
\newblock Place: England.

\bibitem{Enderling2019}
Heiko Enderling, Juan Carlos~López Alfonso, Eduardo Moros, Jimmy~J Caudell,
  and Louis~B Harrison.
\newblock Integrating mathematical modeling into the roadmap for personalized
  adaptive radiation therapy.
\newblock {\em Trends in Cancer}, 5(8):467--474, 2019.

\bibitem{RN16}
A.~Enriquez, S.~Libring, T.~C. Field, J.~Jimenez, T.~Lee, H.~Park, D.~Satoski,
  M.~K. Wendt, S.~Calve, A.~B. Tepole, L.~Solorio, and H.~Lee.
\newblock High-throughput magnetic actuation platform for evaluating the effect
  of mechanical force on 3d tumor microenvironment.
\newblock {\em Adv Funct Mater}, 31(1), 2021.

\bibitem{eykholt2018robust}
Kevin Eykholt, Ivan Evtimov, Earlence Fernandes, Bo~Li, Amir Rahmati, Chaowei
  Xiao, Atul Prakash, Tadayoshi Kohno, and Dawn Song.
\newblock Robust physical-world attacks on deep learning visual classification.
\newblock In {\em Proceedings of the IEEE conference on computer vision and
  pattern recognition}, pages 1625--1634, 2018.

\bibitem{her2ai}
Food and Drug Administration.
\newblock List of cleared or approved companion diagnostic devices (in vitro
  and imaging tools), 2021.

\bibitem{Fritz2021}
Marvin Fritz, Prashant~K. Jha, Tobias K\"{o}ppl, J.~Tinsley Oden, and Barbara
  Wohlmuth.
\newblock Analysis of a new multispecies tumor growth model coupling 3d
  phase-fields with a 1d vascular network.
\newblock {\em Nonlinear Analysis: Real World Applications}, page 103331, 2021.

\bibitem{fu2020deep}
Yabo Fu, Yang Lei, Tonghe Wang, Walter~J Curran, Tian Liu, and Xiaofeng Yang.
\newblock Deep learning in medical image registration: a review.
\newblock {\em Physics in Medicine \& Biology}, 65(20):20TR01, 2020.

\bibitem{RN57}
N.~Fujioka, Y.~Morimoto, T.~Arai, and M.~Kikuchi.
\newblock Discrimination between normal and malignant human gastric tissues by
  fourier transform infrared spectroscopy.
\newblock {\em Cancer Detect Prev}, 28(1):32--6, 2004.

\bibitem{garcez2023neurosymbolic}
Artur~d’Avila Garcez and Luis~C Lamb.
\newblock Neurosymbolic ai: The 3 rd wave.
\newblock {\em Artificial Intelligence Review}, pages 1--20, 2023.

\bibitem{Gholami2016}
Amir Gholami, Andreas Mang, and George Biros.
\newblock An inverse problem formulation for parameter estimation of a
  reaction--diffusion model of low grade gliomas.
\newblock {\em Journal of mathematical biology}, 72:409--433, 2016.

\bibitem{gichoya2022ai}
Judy~Wawira Gichoya, Imon Banerjee, Ananth~Reddy Bhimireddy, John~L Burns,
  Leo~Anthony Celi, Li-Ching Chen, Ramon Correa, Natalie Dullerud, Marzyeh
  Ghassemi, Shih-Cheng Huang, et~al.
\newblock Ai recognition of patient race in medical imaging: a modelling study.
\newblock {\em The Lancet Digital Health}, 4(6):e406--e414, 2022.

\bibitem{gregor2015draw}
Karol Gregor, Ivo Danihelka, Alex Graves, Danilo Rezende, and Daan Wierstra.
\newblock Draw: A recurrent neural network for image generation.
\newblock In {\em International conference on machine learning}, pages
  1462--1471. PMLR, 2015.

\bibitem{Hadjicharalambous2022}
Myrianthi Hadjicharalambous, Eleftherios Ioannou, Nicolas Aristokleous, Kristaq
  Gazeli, Charalambos Anastassiou, and Vasileios Vavourakis.
\newblock Combined anti-angiogenic and cytotoxic treatment of a solid tumour:
  in silico investigation of a xenograft animal model’s digital twin.
\newblock {\em Journal of Theoretical Biology}, 553:111246, 2022.

\bibitem{Han2020}
W.~Han, L.~Qin, C.~Bay, X.~Chen, K.~H. Yu, N.~Miskin, A.~Li, X.~Xu, and
  G.~Young.
\newblock {Deep Transfer Learning and Radiomics Feature Prediction of Survival
  of Patients with High-Grade Gliomas}.
\newblock {\em American Journal of Neuroradiology}, 41(1):40--48, jan 2020.

\bibitem{RN10}
L.~A. Hapach, S.~P. Carey, S.~C. Schwager, P.~V. Taufalele, W.~Wang, J.~A.
  Mosier, N.~Ortiz-Otero, T.~J. McArdle, Z.~E. Goldblatt, M.~C. Lampi,
  F.~Bordeleau, J.~R. Marshall, I.~M. Richardson, J.~Li, M.~R. King, and C.~A.
  Reinhart-King.
\newblock Phenotypic heterogeneity and metastasis of breast cancer cells.
\newblock {\em Cancer Res}, 81(13):3649--3663, 2021.

\bibitem{RN4}
A.~Haque, J.~Engel, S.~A. Teichmann, and T.~Lonnberg.
\newblock A practical guide to single-cell rna-sequencing for biomedical
  research and clinical applications.
\newblock {\em Genome Med}, 9(1):75, 2017.

\bibitem{HernandezBoussard2021}
Tina Hernandez-Boussard, Paul Macklin, Emily~J Greenspan, Amy~L Gryshuk, Eric
  Stahlberg, Tanveer Syeda-Mahmood, and Ilya Shmulevich.
\newblock Digital twins for predictive oncology will be a paradigm shift for
  precision cancer care.
\newblock {\em Nature medicine}, 27(12):2065--2066, 2021.

\bibitem{Hormuth2021}
David~A Hormuth, Karine~A Al~Feghali, Andrew~M Elliott, Thomas~E Yankeelov, and
  Caroline Chung.
\newblock Image-based personalization of computational models for predicting
  response of high-grade glioma to chemoradiation.
\newblock {\em Scientific Reports}, 11(1):8520, 2021.

\bibitem{hormuth_opportunities_2022}
David~A. Hormuth, Maguy Farhat, Chase Christenson, Brandon Curl,
  C.~Chad~Quarles, Caroline Chung, and Thomas~E. Yankeelov.
\newblock Opportunities for improving brain cancer treatment outcomes through
  imaging-based mathematical modeling of the delivery of radiotherapy and
  immunotherapy.
\newblock {\em Advanced Drug Delivery Reviews}, 187:114367, 2022.
\newblock Publisher: Elsevier B.V.

\bibitem{Hormuth2019}
David~A Hormuth, Angela~M Jarrett, Xinzeng Feng, and Thomas~E Yankeelov.
\newblock Calibrating a predictive model of tumor growth and angiogenesis with
  quantitative mri.
\newblock {\em Annals of Biomedical Engineering}, 47:1539--1551, 2019.

\bibitem{Hormuth2020}
David~A Hormuth, Angela~M Jarrett, and Thomas~E Yankeelov.
\newblock Forecasting tumor and vasculature response dynamics to radiation
  therapy via image based mathematical modeling.
\newblock {\em Radiation Oncology}, 15:1--14, 2020.

\bibitem{Hormuth2019a}
David~A. Hormuth~II, Anna~G. Sorace, John Virostko, Richard~G. Abramson,
  Zaver~M. Bhujwalla, Pedro Enriquez-Navas, Robert Gillies, John~D. Hazle,
  Ralph~P. Mason, C.~Chad Quarles, Jared~A. Weis, Jennifer~G. Whisenant,
  Junzhong Xu, and Thomas~E. Yankeelov.
\newblock Translating preclinical {MRI} methods to clinical oncology.
\newblock {\em Journal of Magnetic Resonance Imaging}, 50(5):1377--1392, 2019.

\bibitem{howland2023cellular}
Kennedy~K Howland and Amy Brock.
\newblock Cellular barcoding tracks heterogeneous clones through selective
  pressures and phenotypic transitions.
\newblock {\em Trends in Cancer}, 2023.

\bibitem{IzurzunArana2018}
Itziar Irurzun-Arana, Alvaro Janda, Sergio Ardanza-Trevijano, and I\~{n}aki~F.
  Troc\'{o}niz.
\newblock Optimal dynamic control approach in a multi-objective therapeutic
  scenario: Application to drug delivery in the treatment of prostate cancer.
\newblock {\em PLOS Computational Biology}, 14(4):e1006087, 2018.

\bibitem{isensee2021nnu}
Fabian Isensee, Paul~F Jaeger, Simon~AA Kohl, Jens Petersen, and Klaus~H
  Maier-Hein.
\newblock nnu-net: a self-configuring method for deep learning-based biomedical
  image segmentation.
\newblock {\em Nature methods}, 18(2):203--211, 2021.

\bibitem{Iyengar2012}
Ravi Iyengar, Shan Zhao, Seung-Wook Chung, Donald~E. Mager, and James~M. Gallo.
\newblock Merging systems biology with pharmacodynamics.
\newblock {\em Science Translational Medicine}, 4(126):126ps7, 2012.

\bibitem{jaffray2012image}
David~A Jaffray.
\newblock Image-guided radiotherapy: from current concept to future
  perspectives.
\newblock {\em Nature reviews Clinical oncology}, 9(12):688--699, 2012.

\bibitem{Jarrett2020}
Angela~M. Jarrett, Danial Faghihi, David~A. Hormuth, Ernesto A. B.~F. Lima,
  John Virostko, George Biros, Debra Patt, and Thomas~E. Yankeelov.
\newblock Optimal control theory for personalized therapeutic regimens in
  oncology: Background, history, challenges, and opportunities.
\newblock {\em Journal of Clinical Medicine}, 9(5):1314, 2020.

\bibitem{jarrett_quantitative_2021}
Angela~M. Jarrett, Anum~S. Kazerouni, Chengyue Wu, John Virostko, Anna~G.
  Sorace, Julie~C. {DiCarlo}, David~A. Hormuth, David~A. Ekrut, Debra Patt,
  Boone Goodgame, Sarah Avery, and Thomas~E. Yankeelov.
\newblock Quantitative magnetic resonance imaging and tumor forecasting of
  breast cancer patients in the community setting.
\newblock {\em Nature Protocols}, 16(11):5309--5338, 2021.

\bibitem{jin_positron_2022}
Chentao Jin, Xiaoyun Luo, Xiaoyi Li, Rui Zhou, Yan Zhong, Zhoujiao Xu, Chunyi
  Cui, Xiaoqing Xing, Hong Zhang, and Mei Tian.
\newblock Positron emission tomography molecular imaging-based cancer
  phenotyping.
\newblock {\em Cancer}, 128(14):2704--2716, 2022.
\newblock Place: United States.

\bibitem{RN1}
C.~Jolly and P.~Van~Loo.
\newblock Timing somatic events in the evolution of cancer.
\newblock {\em Genome Biol}, 19(1):95, 2018.

\bibitem{RN17}
S.~Jordahl, L.~Solorio, D.~B. Neale, S.~McDermott, J.~H. Jordahl, A.~Fox,
  C.~Dunlay, A.~Xiao, M.~Brown, M.~Wicha, G.~D. Luker, and J.~Lahann.
\newblock Engineered fibrillar fibronectin networks as three-dimensional tissue
  scaffolds.
\newblock {\em Adv Mater}, 31(46):e1904580, 2019.

\bibitem{RN52}
J.~A.~A. Jothi and V.~M.~A. Rajam.
\newblock A survey on automated cancer diagnosis from histopathology images.
\newblock {\em Artif Intell Rev}, 48(1):50, 2017.

\bibitem{Judenhofer2013}
Martin~S. Judenhofer and Simon~R. Cherry.
\newblock Applications for preclinical {PET/MRI}.
\newblock {\em Seminars in Nuclear Medicine}, 43(1):19--29, 2013.

\bibitem{RN11}
B.~H. Jun, T.~Guo, S.~Libring, M.~K. Chanda, J.~S. Paez, A.~Shinde, M.~K.
  Wendt, P.~P. Vlachos, and L.~Solorio.
\newblock Fibronectin-expressing mesenchymal tumor cells promote breast cancer
  metastasis.
\newblock {\em Cancers (Basel)}, 12(9), 2020.

\bibitem{Kapteyn2021}
Michael~G Kapteyn, Jacob~VR Pretorius, and Karen~E Willcox.
\newblock A probabilistic graphical model foundation for enabling predictive
  digital twins at scale.
\newblock {\em Nature Computational Science}, 1(5):337--347, 2021.

\bibitem{Kazerouni2020}
Anum~S. Kazerouni, Manasa Gadde, Andrea Gardner, David~A. Hormuth, Angela~M.
  Jarrett, Kaitlyn~E. Johnson, Ernesto A.B.~F. Lima, Guillermo Lorenzo, Caleb
  Phillips, Amy Brock, and Thomas~E. Yankeelov.
\newblock Integrating quantitative assays with biologically based mathematical
  modeling for predictive oncology.
\newblock {\em iScience}, 23(12):101807, 2020.

\bibitem{khan2022transformers}
Salman Khan, Muzammal Naseer, Munawar Hayat, Syed~Waqas Zamir, Fahad~Shahbaz
  Khan, and Mubarak Shah.
\newblock Transformers in vision: A survey.
\newblock {\em ACM computing surveys (CSUR)}, 54(10s):1--41, 2022.

\bibitem{RN54}
M.~Khened, A.~Kori, H.~Rajkumar, G.~Krishnamurthi, and B.~Srinivasan.
\newblock A generalized deep learning framework for whole-slide image
  segmentation and analysis.
\newblock {\em Sci Rep}, 11(1):11579, 2021.

\bibitem{Kremheller2019}
Johannes Kremheller, Anh-Tu Vuong, Bernhard~A. Schrefler, and Wolfgang~A. Wall.
\newblock An approach for vascular tumor growth based on a hybrid
  embedded/homogenized treatment of the vasculature within a multiphase porous
  medium model.
\newblock {\em International Journal for Numerical Methods in Biomedical
  Engineering}, 35(11):e3253, 2019.

\bibitem{kuenzi2020predicting}
Brent~M Kuenzi, Jisoo Park, Samson~H Fong, Kyle~S Sanchez, John Lee, Jason~F
  Kreisberg, Jianzhu Ma, and Trey Ideker.
\newblock Predicting drug response and synergy using a deep learning model of
  human cancer cells.
\newblock {\em Cancer cell}, 38(5):672--684, 2020.

\bibitem{lagergren2020biologically}
John~H Lagergren, John~T Nardini, Ruth~E Baker, Matthew~J Simpson, and Kevin~B
  Flores.
\newblock Biologically-informed neural networks guide mechanistic modeling from
  sparse experimental data.
\newblock {\em PLoS computational biology}, 16(12):e1008462, 2020.

\bibitem{Lal2019}
Rayhan~A Lal, Laya Ekhlaspour, Korey Hood, and Bruce Buckingham.
\newblock {Realizing a Closed-Loop (Artificial Pancreas) System for the
  Treatment of Type 1 Diabetes}.
\newblock {\em Endocrine Reviews}, 40(6):1521--1546, 2019.

\bibitem{lemay2022improving}
Andreanne Lemay, Katharina Hoebel, Christopher~P Bridge, Brian Befano, Silvia
  De~Sanjos{\'e}, Didem Egemen, Ana~Cecilia Rodriguez, Mark Schiffman,
  John~Peter Campbell, and Jayashree Kalpathy-Cramer.
\newblock Improving the repeatability of deep learning models with monte carlo
  dropout.
\newblock {\em npj Digital Medicine}, 5(1):174, 2022.

\bibitem{Lenhart2007}
Suzanne Lenhart and John~T Workman.
\newblock {\em Optimal control applied to biological models}.
\newblock Chapman \& Hall/CRC, Philadelphia, PA, 2007.

\bibitem{RN56}
P.~D. Lewis, K.~E. Lewis, R.~Ghosal, S.~Bayliss, A.~J. Lloyd, J.~Wills,
  R.~Godfrey, P.~Kloer, and L.~A. Mur.
\newblock Evaluation of ftir spectroscopy as a diagnostic tool for lung cancer
  using sputum.
\newblock {\em BMC Cancer}, 10:640, 2010.

\bibitem{Lima2017}
E.A.B.F. Lima, J.T. Oden, B.~Wohlmuth, A.~Shahmoradi, D.A. Hormuth, T.E.
  Yankeelov, L.~Scarabosio, and T.~Horger.
\newblock Selection and validation of predictive models of radiation effects on
  tumor growth based on noninvasive imaging data.
\newblock {\em Computer Methods in Applied Mechanics and Engineering},
  327:277--305, 2017.

\bibitem{Lima2022}
Ernesto~A.B.F. Lima, Reid~A.F. Wyde, Anna~G. Sorace, and Thomas~E. Yankeelov.
\newblock Optimizing combination therapy in a murine model of {HER2+} breast
  cancer.
\newblock {\em Computer Methods in Applied Mechanics and Engineering},
  402:115484, 2022.

\bibitem{linardatos2020explainable}
Pantelis Linardatos, Vasilis Papastefanopoulos, and Sotiris Kotsiantis.
\newblock Explainable ai: A review of machine learning interpretability
  methods.
\newblock {\em Entropy}, 23(1):18, 2020.

\bibitem{Lipkova2019}
Jana Lipkov\'{á}, Panagiotis Angelikopoulos, Stephen Wu, Esther Alberts,
  Benedikt Wiestler, Christian Diehl, Christine Preibisch, Thomas Pyka,
  Stephanie~E. Combs, Panagiotis Hadjidoukas, Koen Van~Leemput, Petros
  Koumoutsakos, John Lowengrub, and Bjoern Menze.
\newblock Personalized radiotherapy design for glioblastoma: Integrating
  mathematical tumor models, multimodal scans, and bayesian inference.
\newblock {\em IEEE Transactions on Medical Imaging}, 38(8):1875--1884, 2019.

\bibitem{lone2022liquid}
Saife~N Lone, Sabah Nisar, Tariq Masoodi, Mayank Singh, Arshi Rizwan, Sheema
  Hashem, Wael El-Rifai, Davide Bedognetti, Surinder~K Batra, Mohammad Haris,
  et~al.
\newblock Liquid biopsy: A step closer to transform diagnosis, prognosis and
  future of cancer treatments.
\newblock {\em Molecular cancer}, 21(1):1--22, 2022.

\bibitem{Lorenzo2022}
Guillermo Lorenzo, Nadia {di Muzio}, Chiara~Lucrezia Deantoni, Cesare
  Cozzarini, Andrei Fodor, Alberto Briganti, Francesco Montorsi, V\'{i}ctor~M.
  P\'{e}rez-Garc\'{i}a, Hector Gomez, and Alessandro Reali.
\newblock Patient-specific forecasting of postradiotherapy prostate-specific
  antigen kinetics enables early prediction of biochemical relapse.
\newblock {\em iScience}, 25(11):105430, 2022.

\bibitem{Lorenzo2022b}
Guillermo Lorenzo, Jon~S. Heiselman, Michael~A. Liss, Michael~I. Miga, Hector
  Gomez, Thomas~E. Yankeelov, Thomas~J. Hughes, and Alessandro Reali.
\newblock {Abstract 5064: Patient-specific forecasting of prostate cancer
  growth during active surveillance using an imaging-informed mechanistic
  model}.
\newblock {\em Cancer Research}, 82(12 Supplement):5064, 06 2022.

\bibitem{Lorenzo2023}
Guillermo Lorenzo, David~A Hormuth~II, Angela~M Jarrett, Ernesto~ABF Lima,
  Shashank Subramanian, George Biros, J~Tinsley Oden, Thomas~JR Hughes, and
  Thomas~E Yankeelov.
\newblock Quantitative in vivo imaging to enable tumor forecasting and
  treatment optimization.
\newblock In Igor Balaz and Andrew Adamatzky, editors, {\em Cancer, Complexity,
  Computation}. Springer, 2022.

\bibitem{Lorenzo2019}
Guillermo Lorenzo, Thomas J.~R. Hughes, Pablo Dominguez-Frojan, Alessandro
  Reali, and Hector Gomez.
\newblock Computer simulations suggest that prostate enlargement due to benign
  prostatic hyperplasia mechanically impedes prostate cancer growth.
\newblock {\em Proceedings of the National Academy of Sciences},
  116(4):1152--1161, 2019.

\bibitem{RN18}
K.~M. Lugo-Cintron, M.~M. Gong, J.~M. Ayuso, L.~A. Tomko, D.~J. Beebe,
  M.~Virumbrales-Munoz, and S.~M. Ponik.
\newblock Breast fibroblasts and ecm components modulate breast cancer cell
  migration through the secretion of mmps in a 3d microfluidic co-culture
  model.
\newblock {\em Cancers (Basel)}, 12(5), 2020.

\bibitem{Malone2010}
Hani~R Malone, Omar~N Syed, Michael~S Downes, Anthony~L D'Ambrosio, Donald~O
  Quest, and Michael~G Kaiser.
\newblock Simulation in neurosurgery: a review of computer-based simulation
  environments and their surgical applications.
\newblock {\em Neurosurgery}, 67(4):1105--1116, 2010.

\bibitem{RN53}
N.~Marini, S.~Marchesin, S.~Otalora, M.~Wodzinski, A.~Caputo, M.~van Rijthoven,
  W.~Aswolinskiy, J.~M. Bokhorst, D.~Podareanu, E.~Petters, S.~Boytcheva,
  G.~Buttafuoco, S.~Vatrano, F.~Fraggetta, J.~van~der Laak, M.~Agosti,
  F.~Ciompi, G.~Silvello, H.~Muller, and M.~Atzori.
\newblock Unleashing the potential of digital pathology data by training
  computer-aided diagnosis models without human annotations.
\newblock {\em NPJ Digit Med}, 5(1):102, 2022.

\bibitem{Mascheroni2021}
Pietro Mascheroni, Symeon Savvopoulos, Juan Carlos~L{\'o}pez Alfonso, Michael
  Meyer-Hermann, and Haralampos Hatzikirou.
\newblock Improving personalized tumor growth predictions using a bayesian
  combination of mechanistic modeling and machine learning.
\newblock {\em Communications Medicine}, 1(1):19, 2021.

\bibitem{Menze2015}
Bjoern~H Menze, Andras Jakab, Stefan Bauer, Jayashree Kalpathy-Cramer, Keyvan
  Farahani, Justin Kirby, Yuliya Burren, Nicole Porz, Johannes Slotboom, Roland
  Wiest, et~al.
\newblock The multimodal brain tumor image segmentation benchmark (brats).
\newblock {\em IEEE transactions on medical imaging}, 34(10):1993--2024, 2014.

\bibitem{Metzcar2019}
John Metzcar, Yafei Wang, Randy Heiland, and Paul Macklin.
\newblock A review of cell-based computational modeling in cancer biology.
\newblock {\em JCO clinical cancer informatics}, 2:1--13, 2019.

\bibitem{Nardini2020}
John~T Nardini, John~H Lagergren, Andrea Hawkins-Daarud, Lee Curtin, Bethan
  Morris, Erica~M Rutter, Kristin~R Swanson, and Kevin~B Flores.
\newblock Learning equations from biological data with limited time samples.
\newblock {\em Bulletin of mathematical biology}, 82:1--33, 2020.

\bibitem{neyman1939new}
Jerzy Neyman.
\newblock On a new class of" contagious" distributions, applicable in
  entomology and bacteriology.
\newblock {\em The Annals of Mathematical Statistics}, 10(1):35--57, 1939.

\bibitem{Niederer2021}
Steven~A Niederer, Michael~S Sacks, Mark Girolami, and Karen Willcox.
\newblock Scaling digital twins from the artisanal to the industrial.
\newblock {\em Nature Computational Science}, 1(5):313--320, 2021.

\bibitem{obermeyer2019dissecting}
Ziad Obermeyer, Brian Powers, Christine Vogeli, and Sendhil Mullainathan.
\newblock Dissecting racial bias in an algorithm used to manage the health of
  populations.
\newblock {\em Science}, 366(6464):447--453, 2019.

\bibitem{ocana-tienda_growth_2022}
Beatriz Ocaña-Tienda, Julián Pérez-Beteta, David Molina-García, Beatriz
  Asenjo, Ana Ortiz~de Mendivil, David Albillo, Luís~A Pérez-Romasanta,
  Elisabeth González~del Portillo, Manuel Llorente, Natalia Carballo,
  Estanislao Arana, and Víctor~M Pérez-García.
\newblock Growth dynamics of brain metastases differentiate radiation necrosis
  from recurrence.
\newblock {\em Neuro-Oncology Advances}, 5(1), 2022.
\newblock \_eprint:
  https://academic.oup.com/noa/article-pdf/5/1/vdac179/48957608/vdac179.pdf.

\bibitem{oconnor_dynamic_2011}
J~P~B O'Connor, P~S Tofts, K~A Miles, L~M Parkes, G~Thompson, and A~Jackson.
\newblock Dynamic contrast-enhanced imaging techniques: {CT} and {MRI}.
\newblock {\em The British Journal of Radiology}, 84:S112--S120, 2011.
\newblock Publisher: The British Institute of Radiology. Place: 36 Portland
  Place, London, W1B 1AT.

\bibitem{RN5}
S.~Ogbeide, F.~Giannese, L.~Mincarelli, and I.~C. Macaulay.
\newblock Into the multiverse: advances in single-cell multiomic profiling.
\newblock {\em Trends Genet}, 38(8):831--843, 2022.

\bibitem{RN51}
S.~P. Oliveira, P.~C. Neto, J.~Fraga, D.~Montezuma, A.~Monteiro, J.~Monteiro,
  L.~Ribeiro, S.~Goncalves, I.~M. Pinto, and J.~S. Cardoso.
\newblock Cad systems for colorectal cancer from wsi are still not ready for
  clinical acceptance.
\newblock {\em Sci Rep}, 11(1):14358, 2021.

\bibitem{padhani_diffusion-weighted_2009}
Anwar~R Padhani, Guoying Liu, Dow Mu-Koh, Thomas~L Chenevert, Harriet~C Thoeny,
  Taro Takahara, Andrew Dzik-Jurasz, Brian~D Ross, Marc Van~Cauteren, David
  Collins, Dima~A Hammoud, Gordon J~S Rustin, Bachir Taouli, and Peter~L
  Choyke.
\newblock Diffusion-weighted magnetic resonance imaging as a cancer biomarker:
  Consensus and recommendations.
\newblock {\em Neoplasia (New York, N.Y.)}, 11(2):102--125, 2009.

\bibitem{patel2021opportunities}
Jay Patel, Ken Chang, Syed~Rakin Ahmed, Ikbeom Jang, and Jayashree
  Kalpathy-Cramer.
\newblock Opportunities and challenges for deep learning in brain lesions.
\newblock In {\em International MICCAI Brainlesion Workshop}, pages 25--36.
  Springer, 2021.

\bibitem{peng2022deep}
Jian Peng, Daniel~D Kim, Jay~B Patel, Xiaowei Zeng, Jiaer Huang, Ken Chang,
  Xinping Xun, Chen Zhang, John Sollee, Jing Wu, et~al.
\newblock Deep learning-based automatic tumor burden assessment of pediatric
  high-grade gliomas, medulloblastomas, and other leptomeningeal seeding
  tumors.
\newblock {\em Neuro-oncology}, 24(2):289--299, 2022.

\bibitem{RN55}
S.~Perincheri, A.~W. Levi, R.~Celli, P.~Gershkovich, D.~Rimm, J.~S. Morrow,
  B.~Rothrock, P.~Raciti, D.~Klimstra, and J.~Sinard.
\newblock An independent assessment of an artificial intelligence system for
  prostate cancer detection shows strong diagnostic accuracy.
\newblock {\em Mod Pathol}, 34(8):1588--1595, 2021.

\bibitem{RN20}
P.~P. Provenzano, K.~W. Eliceiri, J.~M. Campbell, D.~R. Inman, J.~G. White, and
  P.~J. Keely.
\newblock Collagen reorganization at the tumor-stromal interface facilitates
  local invasion.
\newblock {\em BMC Med}, 4(1):38, 2006.

\bibitem{quarles_imaging_2019}
C~Chad Quarles, Laura~C Bell, and Ashley~M Stokes.
\newblock Imaging vascular and hemodynamic features of the brain using dynamic
  susceptibility contrast and dynamic contrast enhanced {MRI}.
\newblock {\em {NeuroImage}}, 187:32--55, 2019.

\bibitem{raciti2022clinical}
Patricia Raciti, Jillian Sue, Juan~A Retamero, Rodrigo Ceballos, Ran Godrich,
  Jeremy~D Kunz, Adam Casson, Dilip Thiagarajan, Zahra Ebrahimzadeh, Julian
  Viret, et~al.
\newblock Clinical validation of artificial intelligence--augmented pathology
  diagnosis demonstrates significant gains in diagnostic accuracy in prostate
  cancer detection.
\newblock {\em Archives of Pathology \& Laboratory Medicine}, 2022.

\bibitem{raissi2019physics}
Maziar Raissi, Paris Perdikaris, and George~E Karniadakis.
\newblock Physics-informed neural networks: A deep learning framework for
  solving forward and inverse problems involving nonlinear partial differential
  equations.
\newblock {\em Journal of Computational physics}, 378:686--707, 2019.

\bibitem{rajendran_hypoxia_2004}
J.~G. Rajendran.
\newblock Hypoxia and glucose metabolism in malignant tumors: Evaluation by
  [18f]fluoromisonidazole and [18f]fluorodeoxyglucose positron emission
  tomography imaging.
\newblock {\em Clinical Cancer Research}, 10(7):2245--2252, 2004.

\bibitem{Rasheed2020}
Adil Rasheed, Omer San, and Trond Kvamsdal.
\newblock Digital twin: Values, challenges and enablers from a modeling
  perspective.
\newblock {\em IEEE Access}, 8:21980--22012, 2020.

\bibitem{Rockne2019}
Russell~C Rockne, Andrea Hawkins-Daarud, Kristin~R Swanson, James~P Sluka,
  James~A Glazier, Paul Macklin, David~A Hormuth, Angela~M Jarrett, Ernesto A
  B~F Lima, J~Tinsley~Oden, George Biros, Thomas~E Yankeelov, Kit Curtius,
  Ibrahim Al~Bakir, Dominik Wodarz, Natalia Komarova, Luis Aparicio, Mykola
  Bordyuh, Raul Rabadan, Stacey~D Finley, Heiko Enderling, Jimmy Caudell,
  Eduardo~G Moros, Alexander R~A Anderson, Robert~A Gatenby, Artem Kaznatcheev,
  Peter Jeavons, Nikhil Krishnan, Julia Pelesko, Raoul~R Wadhwa, Nara Yoon,
  Daniel Nichol, Andriy Marusyk, Michael Hinczewski, and Jacob~G Scott.
\newblock The 2019 mathematical oncology roadmap.
\newblock {\em Physical Biology}, 16(4):41005, 2019.
\newblock Publisher: {IOP} Publishing.

\bibitem{Rockne2015}
Russell~C. Rockne, Andrew~D. Trister, Joshua Jacobs, Andrea~J. Hawkins-Daarud,
  Maxwell~L. Neal, Kristi Hendrickson, Maciej~M. Mrugala, Jason~K. Rockhill,
  Paul Kinahan, Kenneth~A. Krohn, and Kristin~R. Swanson.
\newblock A patient-specific computational model of hypoxia-modulated radiation
  resistance in glioblastoma using {F-FMISO-PET}.
\newblock {\em Journal of The Royal Society Interface}, 12(103):20141174, 2015.

\bibitem{DBLP:journals/corr/RonnebergerFB15}
Olaf Ronneberger, Philipp Fischer, and Thomas Brox.
\newblock {U-Net: Convolutional Networks for Biomedical Image Segmentation}.
\newblock {\em CoRR}, abs/1505.0, 2015.

\bibitem{Roque2018}
Tha\'{i}s Roque, Laurent Risser, Veerle Kersemans, Sean Smart, Danny Allen,
  Paul Kinchesh, Stuart Gilchrist, Ana~L. Gomes, Julia~A. Schnabel, and
  Michael~A. Chappell.
\newblock A {DCE-MRI} driven 3-d reaction-diffusion model of solid tumor
  growth.
\newblock {\em IEEE Transactions on Medical Imaging}, 37(3):724--732, 2018.

\bibitem{RN3}
A.~Saadatpour, S.~Lai, G.~Guo, and G.~C. Yuan.
\newblock Single-cell analysis in cancer genomics.
\newblock {\em Trends Genet}, 31(10):576--586, 2015.

\bibitem{Schattler2016}
Heinz Sch\"{a}ttler and Urszula Ledzewicz.
\newblock {\em Optimal control for mathematical models of cancer therapies}.
\newblock Springer, New York, NY, 2016.

\bibitem{Shamanna2020}
Paramesh Shamanna, Banshi Saboo, Suresh Damodharan, Jahangir Mohammed, Maluk
  Mohamed, Terrence Poon, Nathan Kleinman, and Mohamed Thajudeen.
\newblock Reducing hba1c in type 2 diabetes using digital twin
  technology-enabled precision nutrition: A retrospective analysis.
\newblock {\em Diabetes Therapy}, 11:2703--2714, 2020.

\bibitem{Shi2017}
Jinjun Shi, Philip~W Kantoff, Richard Wooster, and Omid~C Farokhzad.
\newblock Cancer nanomedicine: progress, challenges and opportunities.
\newblock {\em Nature Reviews Cancer}, 17(1):20--37, 2017.

\bibitem{RN12}
A.~Shinde, S.~Libring, A.~Alpsoy, A.~Abdullah, J.~A. Schaber, L.~Solorio, and
  M.~K. Wendt.
\newblock Autocrine fibronectin inhibits breast cancer metastasis.
\newblock {\em Mol Cancer Res}, 16(10):1579--1589, 2018.

\bibitem{RN19}
A.~Shinde, J.~S. Paez, S.~Libring, K.~Hopkins, L.~Solorio, and M.~K. Wendt.
\newblock Transglutaminase-2 facilitates extracellular vesicle-mediated
  establishment of the metastatic niche.
\newblock {\em Oncogenesis}, 9(2):16, 2020.

\bibitem{shreve_artificial_2022}
Jacob~T Shreve, Sadia~A Khanani, and Tufia~C Haddad.
\newblock Artificial intelligence in oncology: Current capabilities, future
  opportunities, and ethical considerations.
\newblock {\em American Society of Clinical Oncology Educational Book},
  42:842--851, 2022.

\bibitem{RN7}
R.~L. Siegel, K.~D. Miller, and A.~Jemal.
\newblock Cancer statistics, 2017.
\newblock {\em CA Cancer J Clin}, 67(1):7--30, 2017.

\bibitem{siegel_cancer_2023}
Rebecca~L. Siegel, Kimberly~D. Miller, Nikita~Sandeep Wagle, and Ahmedin Jemal.
\newblock Cancer statistics, 2023.
\newblock {\em {CA}: A Cancer Journal for Clinicians}, 73(1):17--48, 2023.
\newblock \_eprint: https://onlinelibrary.wiley.com/doi/pdf/10.3322/caac.21763.

\bibitem{Slavkova2023}
Kalina~P Slavkova, Sahil~H Patel, Zachary Cacini, Anum~S Kazerouni, Andrea~L
  Gardner, Thomas~E Yankeelov, and David~A Hormuth.
\newblock Mathematical modelling of the dynamics of image-informed tumor
  habitats in a murine model of glioma.
\newblock {\em Scientific Reports}, 13(1):2916, 2023.

\bibitem{sobester2008engineering}
Andr{\'a}s Sobester, Alexander Forrester, and Andy Keane.
\newblock {\em Engineering design via surrogate modelling: a practical guide}.
\newblock John Wiley \& Sons, 2008.

\bibitem{Strobl2021}
Maximilian~A.R. Strobl, Jeffrey West, Yannick Viossat, Mehdi Damaghi, Mark
  Robertson-Tessi, Joel~S. Brown, Robert~A. Gatenby, Philip~K. Maini, and
  Alexander~R.A. Anderson.
\newblock {Turnover Modulates the Need for a Cost of Resistance in Adaptive
  Therapy}.
\newblock {\em Cancer Research}, 81(4):1135--1147, 2021.

\bibitem{Stylianopoulos2013}
Triantafyllos Stylianopoulos, John~D. Martin, Matija Snuderl, Fotios Mpekris,
  Saloni~R. Jain, and Rakesh~K. Jain.
\newblock {Coevolution of Solid Stress and Interstitial Fluid Pressure in
  Tumors During Progression: Implications for Vascular Collapse}.
\newblock {\em Cancer Research}, 73(13):3833--3841, 2013.

\bibitem{Swanson2011}
Kristin~R. Swanson, Russell~C. Rockne, Jonathan Claridge, Mark~A. Chaplain,
  Jr~Alvord, Ellsworth~C., and Alexander~R.A. Anderson.
\newblock {Quantifying the Role of Angiogenesis in Malignant Progression of
  Gliomas: In Silico Modeling Integrates Imaging and Histology}.
\newblock {\em Cancer Research}, 71(24):7366--7375, 2011.

\bibitem{Tao2019}
Fei Tao, He~Zhang, Ang Liu, and A.~Y.~C. Nee.
\newblock Digital twin in industry: State-of-the-art.
\newblock {\em IEEE Transactions on Industrial Informatics}, 15(4):2405--2415,
  2019.

\bibitem{tay2022efficient}
Yi~Tay, Mostafa Dehghani, Dara Bahri, and Donald Metzler.
\newblock Efficient transformers: A survey.
\newblock {\em ACM Computing Surveys}, 55(6):1--28, 2022.

\bibitem{Urcun2021}
St\'{e}phane Urcun, Pierre-Yves Rohan, Wafa Skalli, Pierre Nassoy, St\'{e}phane
  P.~A. Bordas, and Giuseppe Scium\`{e}.
\newblock Digital twinning of cellular capsule technology: Emerging outcomes
  from the perspective of porous media mechanics.
\newblock {\em PLOS ONE}, 16(7):e0254512, 2021.

\bibitem{vandenberghe2017relevance}
Michel~E Vandenberghe, Marietta~LJ Scott, Paul~W Scorer, Magnus S{\"o}derberg,
  Denis Balcerzak, and Craig Barker.
\newblock Relevance of deep learning to facilitate the diagnosis of her2 status
  in breast cancer.
\newblock {\em Scientific reports}, 7(1):1--11, 2017.

\bibitem{vaupel_treatment_2001}
Peter Vaupel, Oliver Thews, and Michael Hoeckel.
\newblock Treatment resistance of solid tumors.
\newblock {\em Medical Oncology}, 18(4):243--259, 2001.

\bibitem{Vavourakis2018}
Vasileios Vavourakis, Triantafyllos Stylianopoulos, and Peter~A. Wijeratne.
\newblock In-silico dynamic analysis of cytotoxic drug administration to solid
  tumours: Effect of binding affinity and vessel permeability.
\newblock {\em PLOS Computational Biology}, 14(10):e1006460, 2018.

\bibitem{RN14}
S.~M. Venis, H.~R. Moon, Y.~Yang, S.~M. Utturkar, S.~F. Konieczny, and B.~Han.
\newblock Engineering of a functional pancreatic acinus with reprogrammed
  cancer cells by induced ptf1a expression.
\newblock {\em Lab Chip}, 21(19):3675--3685, 2021.

\bibitem{Viguerie2022}
Alex Viguerie, Malú Grave, Gabriel~F. Barros, Guillermo Lorenzo, Alessandro
  Reali, and Alvaro L. G.~A. Coutinho.
\newblock {Data-Driven Simulation of Fisher–Kolmogorov Tumor Growth Models
  Using Dynamic Mode Decomposition}.
\newblock {\em Journal of Biomechanical Engineering}, 144(12), 2022.
\newblock 121001.

\bibitem{vinuesa2021interpretable}
Ricardo Vinuesa and Beril Sirmacek.
\newblock Interpretable deep-learning models to help achieve the sustainable
  development goals.
\newblock {\em Nature Machine Intelligence}, 3(11):926--926, 2021.

\bibitem{RN15}
M.~Virumbrales-Munoz, J.~M. Ayuso, J.~R. Loken, K.~M. Denecke, S.~Rehman, M.~C.
  Skala, E.~J. Abel, and D.~J. Beebe.
\newblock Microphysiological model of renal cell carcinoma to inform
  anti-angiogenic therapy.
\newblock {\em Biomaterials}, 283:121454, 2022.

\bibitem{RN6}
L.~F. Vistain and S.~Tay.
\newblock Single-cell proteomics.
\newblock {\em Trends Biochem Sci}, 46(8):661--672, 2021.

\bibitem{wagner2021radiomics}
Matthias~W Wagner, Khashayar Namdar, Asthik Biswas, Suranna Monah, Farzad
  Khalvati, and Birgit~B Ertl-Wagner.
\newblock Radiomics, machine learning, and artificial intelligence—what the
  neuroradiologist needs to know.
\newblock {\em Neuroradiology}, pages 1--11, 2021.

\bibitem{RN13}
D.~R. Welch and D.~R. Hurst.
\newblock Defining the hallmarks of metastasis.
\newblock {\em Cancer Res}, 79(12):3011--3027, 2019.

\bibitem{WembacherSchroeder2021}
Eva Wembacher-Schroeder, Nicole Kerstein, Evan~D. Bander, Neeta Pandit-Taskar,
  Rowena Thomson, and Mark~M. Souweidane.
\newblock Evaluation of a patient-specific algorithm for predicting
  distribution for convection-enhanced drug delivery into the brainstem of
  patients with diffuse intrinsic pontine glioma.
\newblock {\em Journal of Neurosurgery: Pediatrics}, 28(1):34 -- 42, 2021.

\bibitem{wen_updated_2010}
Patrick~Y Wen, David~R Macdonald, David~A Reardon, Timothy~F Cloughesy,
  A~Gregory Sorensen, Evanthia Galanis, John {DeGroot}, Wolfgang Wick, Mark~R
  Gilbert, Andrew~B Lassman, Christina Tsien, Tom Mikkelsen, Eric~T Wong,
  Marc~C Chamberlain, Roger Stupp, Kathleen~R Lamborn, Michael~A Vogelbaum,
  Martin~J van~den Bent, and Susan~M Chang.
\newblock Updated response assessment criteria for high-grade gliomas: Response
  assessment in neuro-oncology working group.
\newblock {\em Journal of Clinical Oncology}, 28(11):1963--1972, 2010.

\bibitem{RN9}
M.~K. Wendt and W.~P. Schiemann.
\newblock Therapeutic targeting of the focal adhesion complex prevents
  oncogenic tgf-beta signaling and metastasis.
\newblock {\em Breast Cancer Res}, 11(5):R68, 2009.

\bibitem{Wise2008}
S.M. Wise, J.S. Lowengrub, H.B. Frieboes, and V.~Cristini.
\newblock Three-dimensional multispecies nonlinear tumor growth—i: Model and
  numerical method.
\newblock {\em Journal of Theoretical Biology}, 253(3):524--543, 2008.

\bibitem{Wong2016}
Ken C.~L. Wong, Ronald~M. Summers, Electron Kebebew, and Jianhua Yao.
\newblock Pancreatic tumor growth prediction with elastic-growth decomposition,
  image-derived motion, and {FDM-FEM} coupling.
\newblock {\em IEEE Transactions on Medical Imaging}, 36(1):111--123, 2017.

\bibitem{Woodall2021}
Ryan~T Woodall, David~A Hormuth~II, Chengyue Wu, Michael~RA Abdelmalik,
  William~T Phillips, Ande Bao, Thomas~JR Hughes, Andrew~J Brenner, and
  Thomas~E Yankeelov.
\newblock Patient specific, imaging-informed modeling of rhenium-186
  nanoliposome delivery via convection-enhanced delivery in glioblastoma
  multiforme.
\newblock {\em Biomedical Physics \& Engineering Express}, 7(4):045012, 2021.

\bibitem{Wu2022}
Chengyue Wu, David~A. Hormuth, Guillermo Lorenzo, Angela~M. Jarrett, Federico
  Pineda, Frederick~M. Howard, Gregory~S. Karczmar, and Thomas~E. Yankeelov.
\newblock Towards patient-specific optimization of neoadjuvant treatment
  protocols for breast cancer based on image-guided fluid dynamics.
\newblock {\em IEEE Transactions on Biomedical Engineering}, 69(11):3334--3344,
  2022.

\bibitem{Wu2022b}
Chengyue Wu, Angela~M. Jarrett, Zijian Zhou, Nabil Elshafeey, Beatriz~E.
  Adrada, Rosalind~P. Candelaria, Rania~M.M. Mohamed, Medine Boge, Lei Huo,
  Jason~B. White, Debu Tripathy, Vicente Valero, Jennifer~K. Litton, Clinton
  Yam, Jong~Bum Son, Jingfei Ma, Gaiane~M. Rauch, and Thomas~E. Yankeelov.
\newblock {MRI-Based Digital Models Forecast Patient-Specific Treatment
  Responses to Neoadjuvant Chemotherapy in Triple-Negative Breast Cancer}.
\newblock {\em Cancer Research}, 82(18):3394--3404, 2022.

\bibitem{Wu2022a}
Chengyue Wu, Guillermo Lorenzo, II~Hormuth, David~A., Ernesto A. B.~F. Lima,
  Kalina~P. Slavkova, Julie~C. DiCarlo, John Virostko, Caleb~M. Phillips, Debra
  Patt, Caroline Chung, and Thomas~E. Yankeelov.
\newblock {Integrating mechanism-based modeling with biomedical imaging to
  build practical digital twins for clinical oncology}.
\newblock {\em Biophysics Reviews}, 3(2):021304, 2022.

\bibitem{RN2}
F.~Wu, J.~Fan, Y.~He, A.~Xiong, J.~Yu, Y.~Li, Y.~Zhang, W.~Zhao, F.~Zhou,
  W.~Li, J.~Zhang, X.~Zhang, M.~Qiao, G.~Gao, S.~Chen, X.~Chen, X.~Li, L.~Hou,
  C.~Wu, C.~Su, S.~Ren, M.~Odenthal, R.~Buettner, N.~Fang, and C.~Zhou.
\newblock Single-cell profiling of tumor heterogeneity and the microenvironment
  in advanced non-small cell lung cancer.
\newblock {\em Nat Commun}, 12(1):2540, 2021.

\bibitem{Xu2020}
Jiangping Xu, Guillermo Vilanova, and Hector Gomez.
\newblock Phase-field model of vascular tumor growth: Three-dimensional
  geometry of the vascular network and integration with imaging data.
\newblock {\em Computer Methods in Applied Mechanics and Engineering},
  359:112648, 2020.

\bibitem{xu2019deep}
Yiwen Xu, Ahmed Hosny, Roman Zeleznik, Chintan Parmar, Thibaud Coroller, Idalid
  Franco, Raymond~H Mak, and Hugo~JWL Aerts.
\newblock Deep learning predicts lung cancer treatment response from serial
  medical imaging.
\newblock {\em Clinical Cancer Research}, 25(11):3266--3275, 2019.

\bibitem{Yang2022}
Emily~Y. Yang, Grant~R. Howard, Amy Brock, Thomas~E. Yankeelov, and Guillermo
  Lorenzo.
\newblock Mathematical characterization of population dynamics in breast cancer
  cells treated with doxorubicin.
\newblock {\em Frontiers in Molecular Biosciences}, 9, 2022.

\bibitem{Yankeelov2016}
Thomas~E Yankeelov, Gary An, Oliver Saut, E~Georg Luebeck, Aleksander~S Popel,
  Benjamin Ribba, Paolo Vicini, Xiaobo Zhou, Jared~A Weis, Kaiming Ye, et~al.
\newblock Multi-scale modeling in clinical oncology: opportunities and barriers
  to success.
\newblock {\em Annals of biomedical engineering}, 44:2626--2641, 2016.

\bibitem{yankeelov_toward_2015}
Thomas~E Yankeelov, Vito Quaranta, Katherine~J Evans, and Erin~C Rericha.
\newblock Toward a science of tumor forecasting for clinical oncology.
\newblock {\em Cancer Research}, 75(6):918--923, 2015.

\bibitem{Yin2019}
Anyue Yin, Dirk Jan~A.R. Moes, Johan~G.C. van Hasselt, Jesse~J. Swen, and
  Henk-Jan Guchelaar.
\newblock A review of mathematical models for tumor dynamics and treatment
  resistance evolution of solid tumors.
\newblock {\em CPT: Pharmacometrics \& Systems Pharmacology}, 8(10):720--737,
  2019.

\bibitem{Zahid2021}
Mohammad~U. Zahid, Nuverah Mohsin, Abdallah~S.R. Mohamed, Jimmy~J. Caudell,
  Louis~B. Harrison, Clifton~D. Fuller, Eduardo~G. Moros, and Heiko Enderling.
\newblock Forecasting individual patient response to radiation therapy in head
  and neck cancer with a dynamic carrying capacity model.
\newblock {\em International Journal of Radiation Oncology*Biology*Physics},
  111(3):693--704, 2021.

\bibitem{zhu2018image}
Bo~Zhu, Jeremiah~Z Liu, Stephen~F Cauley, Bruce~R Rosen, and Matthew~S Rosen.
\newblock Image reconstruction by domain-transform manifold learning.
\newblock {\em Nature}, 555(7697):487--492, 2018.

\end{thebibliography}

\end{document}